\documentclass[journal]{IEEEtran}
\usepackage[dvips]{graphicx}
\graphicspath{{./eps/}}
\DeclareGraphicsExtensions{.eps}
\usepackage{subfigure}
\usepackage[caption=false,font=normalsize,labelfont=sf,textfont=sf]{subfig}
\usepackage{mathrsfs}
\usepackage{times}
\usepackage{bm}
\usepackage{subfigure}
\usepackage{color}
\usepackage{amsfonts}
\usepackage{amssymb}
\usepackage{graphicx}
\usepackage{algorithm}
\usepackage{algorithmic}
\usepackage{color}
\usepackage{multirow}
\usepackage{dsfont}
\usepackage{amsmath}
\usepackage{amsthm}
\usepackage{amsfonts}
\usepackage{amssymb}

\newtheorem{remark}{Remark}

\IEEEoverridecommandlockouts

\usepackage{algorithmic}
\usepackage{array}
\usepackage{threeparttable}
\usepackage{enumerate}
\usepackage{balance}
\begin{document}
	\title{Mean Field Game-based Waveform Precoding Design for Mobile Crowd Integrated Sensing, Communication, and Computation Systems
	}
\author
{\IEEEauthorblockN{Dezhi Wang, Chongwen Huang, \IEEEmembership{Member,~IEEE},
 Jiguang He, \IEEEmembership{Senior Member,~IEEE}, Xiaoming Chen, \IEEEmembership{Senior Member,~IEEE}, Wei Wang, \IEEEmembership{Senior Member,~IEEE}, Zhaoyang Zhang, \IEEEmembership{Senior Member,~IEEE},
 Zhu Han, \IEEEmembership{Fellow,~IEEE}, and M{\'e}rouane  Debbah, \IEEEmembership{Fellow,~IEEE}}\\
\IEEEcompsocitemizethanks{
	\IEEEcompsocthanksitem 
	Part of this work has been presented in Globecom Workshops, Dec. 2023~\cite{wang2023GC}. \protect
	\IEEEcompsocthanksitem 
	D.~Wang and C. Huang are with the College of Information Science and Electronic Engineering, Zhejiang University, Hangzhou 310027, China, with the State Key Laboratory of Integrated Service Networks, Xidian University, Xi’an 710071, China, and Zhejiang Provincial Key Laboratory of Info. Proc., Commun. \& Netw. (IPCAN), Hangzhou 310027, China. (E-mails: dz\_wang@zju.edu.cn; chongwenhuang@zju.edu.cn).
	\IEEEcompsocthanksitem X. Chen, W.~Wang, and Z.~Zhang are with the College of Information Science and Electronic Engineering, and also with Zhejiang Provincial Key Laboratory of Information Processing, Communication and Networking, 
	Zhejiang University, Hangzhou 310027, China (Emails: chen\_xiaoming@zju.edu.cn; wangw@zju.edu.cn; ning\_ming@zju.edu.cn).\protect
	\IEEEcompsocthanksitem J. He is with Technology Innovation Institute, Masdar City 9639, Abu Dhabi, United Arab Emirates (E-mail:jiguang.he@tii.ae).\protect
	% note need leading \protect in 
	\IEEEcompsocthanksitem Z. Han is with the Department of Electrical and Computer Engineering in the University of Houston, Houston, TX 77004 USA (E-mail: hanzhu22@gmail.com).\protect
	\IEEEcompsocthanksitem M. Debbah is with Khalifa University of Science and Technology, P O Box 127788, Abu Dhabi, UAE (E-mail: merouane.debbah@ku.ac.ae).\protect
	}
	
}
\maketitle
\vspace{-2em}
\begin{abstract}

Data collection and processing timely is crucial for mobile crowd integrated sensing, communication, and computation~(ISCC) systems with various applications such as smart home and connected cars, which requires numerous integrated sensing and communication~(ISAC) devices to sense the targets and offload the data to the base station~(BS) for further processing. However, as the number of ISAC devices growing, there exists intensive interactions among ISAC devices in the processes of data collection and processing since they share the common network resources. In this paper, we consider the environment sensing problem in the large-scale mobile crowd ISCC systems and propose an efficient waveform precoding design algorithm based on the mean field game~(MFG). Specifically, to handle the complex interactions among large-scale ISAC devices, we first utilize the MFG method to transform the influence from other ISAC devices into the mean field term and derive the Fokker-Planck-Kolmogorov equation,  which models the evolution of the system state. Then, we derive the cost function based on the mean field term and reformulate the waveform precoding design problem. Next, we utilize the G-prox primal-dual hybrid gradient algorithm to solve the reformulated problem and analyze the computational complexity of the proposed algorithm. Finally, simulation results demonstrate that the proposed algorithm can solve the interactions among large-scale ISAC devices effectively in the ISCC process. In addition, compared with other baselines, the proposed waveform precoding design algorithm has advantages in improving communication performance and reducing cost function.
\end{abstract}

%\newpage
\section{Introduction}
Nowadays, with the rapid development of the Internet of Things~(IoT), numerous applications continue to emerge, such as smart home, autonomous driving,  and pollution monitoring~\cite{jamshed2022challenges}. These applications require a massive amount of data that must be collected by devices and transmitted to servers for additional processing. Mobile crowd sensing~(MCS) has been proposed as an emerging paradigm for collecting data and sensing~\cite{capponi2019survey,zhu2020deep}. It involves outsourcing sensing tasks to ordinary individuals equipped with mobile smart devices such as mobile terminals,  wearable devices, and so on~\cite{zhang2022achieving}. MCS employs devices as the fundamental sensing unit to carry out the sensing tasks through IoT and wireless communication sensor networks~\cite{xiong2020edge}.

To achieve data sensing and processing in MCS systems, each device must have dual capabilities of sensing and communication~\cite{liu2019data}. It is natural to integrate the sensing function into the communication network to improve the spectral efficiency, which is also referred to as integrated sensing and communication~(ISAC)~\cite{luong2021radio}. However, the ISAC system cannot process data timely due to the limited computing resources. With the advancement of multi-access edge computing~(MEC) technology~\cite{wang2023delay}, this issue can be solved by integrating MEC into the MCS system~\cite{he2023integrated,wen2023task2}, which forms the mobile crowd integrated sensing, communication, and computation~(ISCC) system. In this multifunctional system, how to perform cross-layer optimization to complete data collection, transmission, and computation, and improve the performance of the system is the focus of current research.

As an emerging technology, ISAC has received lots of attention in recent years~\cite{liu2022integrated,cui2021integrating}. Du~\emph{et. al}~\cite{du2022integrated}~considered the geometry of vehicles and designed the sensing-aided beamforming for vehicle-to-infrastructure communication by exploiting ISAC functionalities at the roadsides unit. Hua~\emph{et. al}~\cite{hua2023optimal} studied the transmit beamforming in a downlink ISAC system and maximized the radar sensing performance while ensuring the communication metrics. Tong~\emph{et. al}~\cite{tong2021joint} exploited the sparsity of both the structured user signals
and the unstructured environment and proposed an iterative and incremental joint multi-user communication and environment sensing scheme in ISAC systems. Gan~\emph{et. al}~\cite{gan2022near} considered the integration of holographic intelligent surface into a millimeter-wave localization system and obtained the Fisher information matrix and Cram{\'e}r-Rao lower bound.  In addition, with the growing demand for computing applications in future networks, many research focus on combining the ISAC with edge computing~\cite{wen2023task}, which can deal with the application timely. Qi~\emph{et. al}~\cite{qi2022integrating} provided a unified framework ISCC to optimize the limited system resources for 6G wireless networks and derived two typical joint beamforming design algorithms based on multi-objective optimization problems. Ding~\emph{et. al}~\cite{ding2022joint} considered a multi-objective optimization problem in an ISCC architecture and optimized the individual transmit precoding for radar, communication, and computation resource allocation by an iterative optimization algorithm. Li~\emph{et. al}~\cite{li2023integrated} developed an ISCC over-the-air~(ISCCO) framework and designed the separated and shared schemes to support the multiple-input-multiple-output~(MIMO) ISCCO simultaneously. Huang~\emph{et. al}~\cite{huang2022integrated} proposed an ISAC assisted energy-efficient MEC and leveraged advanced intelligent reflecting surface to improve both the performance of the radar sensing and MEC.

 The above research mainly focuses on the applications and resource optimization in ISAC and ISCC systems.  In the future network, the ISCC technology will be widely employed to meet the extreme needs of users, one particular interest area is to apply ISCC to the MCS systems~\cite{li2023quality}. Xiong~\emph{et. al}~\cite{xiong2019task}~proposed a task-oriented user selection incentive mechanism in the MCS with the aid of MEC and maximized the similarity of the task requirements and the preferences of users by designing a secure multi-party sorting protocol. Li~\emph{et.~al}~\cite{li2022joint} proposed an ISCC framework for multi-dimensional resource constrained MCS systems and maximized the number of bits processed in the task. Zhou~\emph{et. al}~\cite{zhou2022joint} considered the sensing-and-transmission energy consumption minimization problem and jointly optimized sensing and transmission rates over the time. Cai \emph{et. al}~\cite{cai2021cooperative} proposed a novel cooperative data sensing and computation offloading scheme based on multi-agent deep reinforcement learning for the UAV-assisted MCS system to maximize the overall system utility. However, the above algorithms may have some limitations in dealing with the mobile crowd ISCC system. In this system, each device should compete against others due to the limited resources~\cite{kang2021task}. The existing algorithm barely considers the completion among devices. On the other hand, large-scale devices need to collect data and transmit the data to the BS for computation, where each device should calculate the influence from all other devices. However, when the number of devices is large, the computational complexity is very high, which is impractical.

The existing research has studied the coupling of sensing and communication functions~\cite{xing2023joint,xiong2023fundamental,an2023fundamental}. However, the coupling among ISAC devices in ISCC process is rarely studied in the above research. In particular, with the continuous growth of ISAC devices in the future MCS system, the computational complexity of existing algorithms will significantly increase due to the rapid increase of interactions among them~\cite{gao2022energyed}. Consequently, it is imperative to develop a suitable algorithm that can handle the interactions among large-scale devices and reduce the computational complexity. In this paper, we consider the waveform precoding design problem in the large-scale mobile crowd ISCC system, and each ISAC device designs the waveform precoding to minimize its energy consumption and computational cost in the ISCC process, where the interactions among ISAC devices should be considered. Specifically, in the sensing and communication processes, each ISAC device should consider the influence from other ISAC devices, such as the influence on signal-interference-to-noise-ratio~(SINR). In addition, when lots of data from ISAC devices is offloaded to edge servers, the delay and cost for processing the data will increase significantly in the computation process. The main challenges are as follows:
%In such a large-scale MCS system, it is challenging to make the distributed beamforming design decision for each sensor since all sensors are coupled together and each sensor cannot obtain the detail information about sensors. Traditional distributed algorithms 

\begin{itemize}
\item \textbf{The intensive interactions among ISAC devices:} In the large-scale mobile crowd ISCC system, each ISAC device completes against others due to the limited communication resources. Therefore, each ISAC device should consider the interactions among ISAC devices when it designs the waveform precoding. However, it is not trivial for each ISAC device to formulate the interactions among ISAC devices in such a large-scale mobile crowd ISCC system, which makes it a challenge.
\item \textbf{The multi-functional coupling joint optimization:} Besides considering the interactions among ISAC devices, it is also necessary to consider the multi-functional coupling in the ISCC process to minimize the cost for each ISAC device. However, the performance of each function is affected by the other two functions since the data collection, transmission, and computation are coupled together. Therefore, resource scheduling and performance balancing among the three functions present significant challenges. 
\end{itemize}

To solve the above issues, we propose an efficient waveform precoding design algorithm based on mean field game~(MFG), which studies the game with large-scale players, where a certain player makes the corresponding decision against the collective behavior of players, i.e., mean field term. In the mean field game, each player has a negligible effect on the overall system. Consequently, a typical player makes an optimal decision by primarily focusing on the mean field term, rather than considering the strategies of all other players~\cite{yang2017mean}, which is independent of the number of ISAC devices, resulting in reduced computational complexity. In the large-scale mobile crowd ISCC system, all ISAC devices are interactive when they design the waveform precoding, which cannot be solved efficiently by using traditional methods. Considering these, we optimize the overall performance of sensing, communication, and computation by simultaneously minimizing the energy consumption and computational cost while satisfying the constraint of sensing accuracy using the MFG method.
The main contributions of this paper are summarized as
\begin{itemize}
	\item To handle the issue of massive interaction among large-scale ISAC devices in the mobile crowd ISCC system, we employ the MFG method to formulate the influence from other ISAC devices into the mean field term, which estimates the collective states of the ISAC devices as a probability distribution. Then, we derive the Fokker-Planck-Kolmogorov~(FPK) equation based on the mean field term, which models the evolution of the system state.
	\item To deal with the problem of cross-layer optimization among multiple functions, we first formulate the problem by minimizing the energy consumption and computational cost under the constraint of sensing accuracy, and reformulate the problem based on the MFG method. Then, we utilize the majorization minimization~(MM) method to transform the sensing accuracy constraint into a convex one, and then leverage the Taylor expansion to transform the FPK equation into a linear form that can be solved through the G-prox primal-dual hybrid gradient (PDHG) algorithm.
	\item Furthermore, we analyze the complexity of the proposed G-prox PDHG algorithm, which is directly proportional to the state space and independent of the number of ISAC devices. In addition, we also validate the performance of the proposed algorithm via the simulation, which shows that the proposed waveform precoding design algorithm can effectively solve the complex interactions among large-scale ISAC devices in the ISCC process, and has advantages in improving communication performance and reducing the cost function.
\end{itemize}

The remainder of this paper is organized as follows. Section~II presents the system model. The problem is formulated in Section III. In Section IV, we reformulate the problem based on the MFG, and in Section V, we give the solution to the waveform precoding design problem. The simulation results are given in Section VI, and finally, we conclude the paper in Section~VII.

\emph{Notations:} To represent a column vector and matrix, bold lowercase, and uppercase symbols are used (e.g., $\bm{x}$ and $\bm{X}$), respectively. For an arbitrary-size matrix $\bm{M}$, $\bm{M}^H$, $\bm{M}^T$, $\bm{M}^{-1}$, and $\text{vec}({\bm{M}})$ denote its conjugate transpose, transpose, inverse, and vectorization operation, respectively.  $\|\cdot\|_2$ presents the $\mathcal{L}_2$ norm of its argument function; $\|\cdot\|_F$ denotes the Frobenius norm of its argument matrix; $\bm{I}_x$ is an $x$-dimensional identity matrix.  The set of real numbers is denoted by $\mathbb{R}$, while the set of the complex numbers is denoted by $\mathbb{C}$. Further, $\bm{x}\sim\mathcal{CN}(0, \bm{\Sigma})$ indicates that $\bm{x}$ is a complex Gaussian vector with zero mean and covariance matrix $\bm{\Sigma}$, and $\triangle$ denotes the Laplace transform.

 \section{System Model}
\begin{figure}
 	{	\centering \includegraphics[width=3.1in]{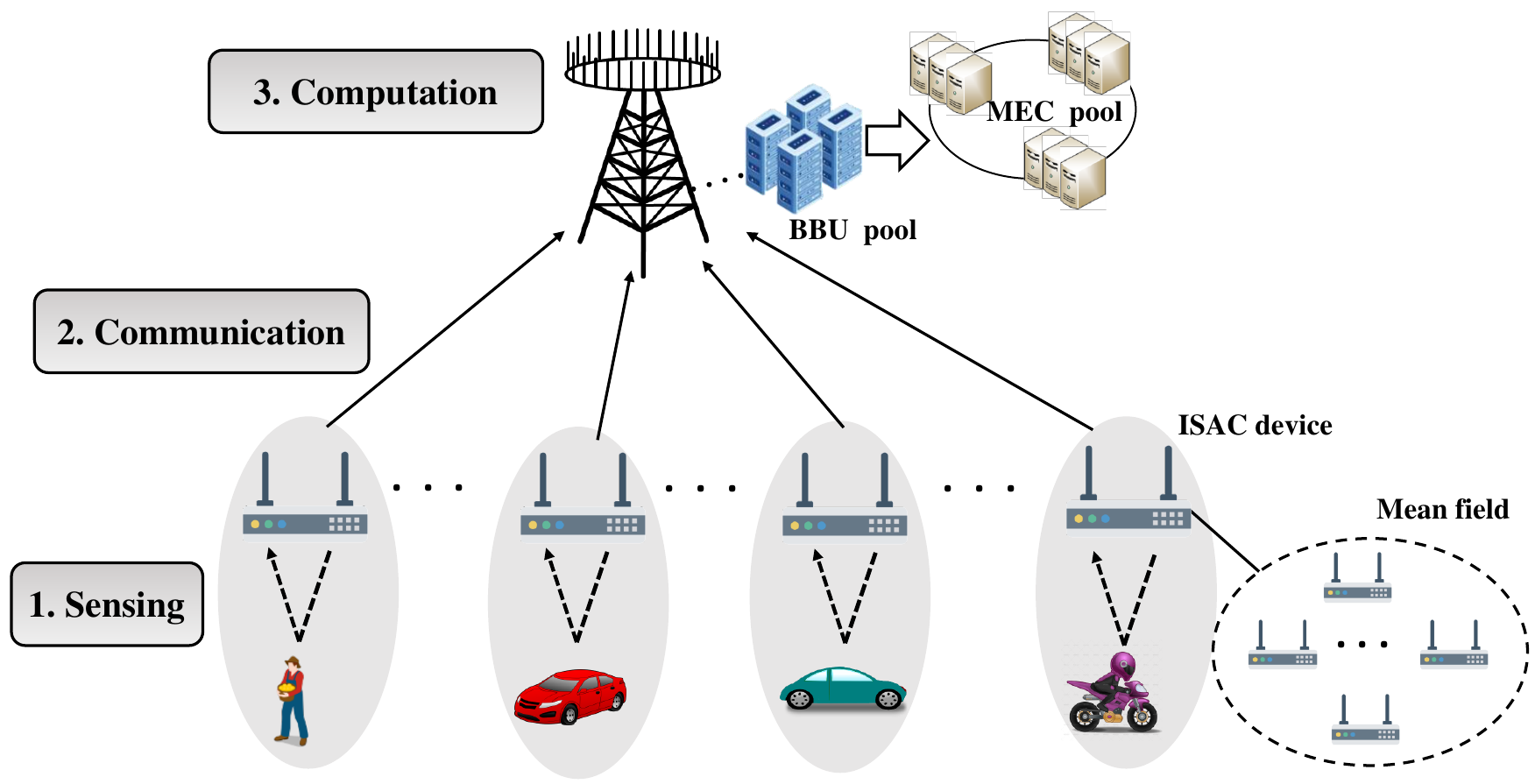}
 		\caption{System model for the mobile crowd ISCC system}
 		\label{fig:system model}
 	}
 \end{figure}

In this paper, we consider a mobile crowd ISCC system, as shown in Fig.~\ref{fig:system model}, which contains one base-station~(BS) with the MEC pool and large-scale ISAC devices, denoted as $\mathcal{N}=\{1,2,\cdots,N\}$. In the system, each ISAC device has $K$ antennas and integrates with communication and sensing functions, and the BS is equipped with $M$ antennas. For the environment sensing task, each ISAC device first performs the sensing task and collects the sensing echo signal. Then, they offload all sensing tasks timely to the BS for processing and obtaining the sensing information via the wireless link. The specific details of the system model are given as follows.
 \subsection{Sensing Model}
  Define $\bm{x}_i(t)$ as the $K\times1$ ISAC signal model for ISAC device $i$ at time slot $t$, which is 
  \begin{eqnarray}
  	\bm{x}_i(t)=\bm{w}_{r,i}(t)s_{r,i}(t)+\bm{w}_{c,i}(t)s_{c,i}(t)=\bm{W}_i\bm{s}_i(t),
  \end{eqnarray}
  where $s_{r,i}(t)\in\mathbb{C}$ is the sensing probing signal for ISAC device $i$ at time slot $t$ with $\mathbb{E}\{s_{r,i}(t)s_{r,i}^H(t)\}$ $=1$, $s_{c,i}(t)\in\mathbb{C}$ is the transmission signal for ISAC device $i$ at time slot $t$ with $\mathbb{E}\{s_{c,i}(t)s_{c,i}^H(t)\}=1$,
  $\bm{w}_{r,i}(t)\in\mathbb{C}^{K\times 1}$ and  $\bm{w}_{c,i}(t)\in\mathbb{C}^{K\times 1}$ are the sensing and communication waveform precoding design vectors for ISAC device $i$  at time slot $t$, respectively. For brevity, define the overall waveform precoding design matrix as $\bm{W}_i(t)\overset{\bigtriangleup}{=}[\bm{w}_{r,i}(t)~ \bm{w}_{c,i}(t)]$ and the overall signal vector as $\bm{s}_i(t)\overset{\bigtriangleup}{=}[s_{r,i}(t)~s_{c,i}(t)]$.
  
  We consider that the sensing target in the far field is located at the direction $\theta_i$ for ISAC device $i$. Therefore, the received signal by ISAC device $i$ at time slot $t$
  is\footnote{In this model, we mainly consider the large-scale fading. For small-scale fading, the channel amplitude and phase are time-varying. In this scenario, we can extend the proposed algorithm into the dynamic channel path loss and phase, since the proposed algorithm is dynamic.}
  \begin{eqnarray}\label{received_radar_signal}
  	&&\bm{y}_{r,i}(t)=\alpha_ie^{-jw_D}\bm{A}_i(\theta_i)\bm{x}_i(t-t_i)\nonumber\\
  	&&~~~~~~+\sum_{j\neq i,j\in\mathcal{N}}\bm{H}_{i,j}(t)\bm{x}_j(t)+\bm{z}_i(t),
  \end{eqnarray}
  where $\alpha_i\in\mathbb{C}$ is the complex path loss between ISAC device $i$ and the target, $w_D$ is the Doppler frequency shift, $t_i$ is the sum of propagation delays, $\bm{H}_{i,j}(t)$ is the interference channel matrix from ISAC device $i$ to ISAC device $j$, $\bm{z}_i$ is the received noise with $\bm{z}_i\sim\mathcal{CN}(0,\sigma_R^2\bm{I}_R)$, and $\bm{A}_i(\theta_i)=\bm{a}_{S,i}(\theta_i)\bm{a}^H_{T,i}(\theta_i)$, where $\bm{a}_{S,i}(\theta_i)\in\mathbb{C}^{K\times1}$ and $\bm{a}_{T,i}(\theta_i)\in\mathbb{C}^{K\times1}$ are the transmit and receive array steering vectors for ISAC device $i$, respectively.
  
  Since the target is far-field, the path loss $\alpha_i$ is assumed to be identical for all transmit and receive antenna elements~\cite{khawar2015target}. In addition, the target moves a small distance over a slot and the sensing process is assumed in a single range-Doppler bin on the sending detector, therefore, the Doppler frequency shift $w_D$ and delay $t_i$ can be completely compensated~\cite{liu2018mimo,bekkerman2006target}. Therefore, we can simplify the received signal in \eqref{received_radar_signal} as 
  \begin{eqnarray}\label{simply_received_radar_signal}
   \bm{y}_{r,i}(t)=\alpha_i\bm{A}_i(\theta_i)\bm{x}_i(t)+\sum_{j\neq i,j\in\mathcal{N}}\bm{H}_{i,j}(t)\bm{x}_j(t)+\bm{z}_i(t).
  \end{eqnarray}
   
  In the sensing process, the transmit and return channels of the signal are the same. Thus, the transmit and receive array steering vector $\bm{a}_{S_i}(\theta_i)$, $\bm{a}_{T_i}(\theta_i)$ are the same, i.e., $\bm{a}_{S,i}(\theta_i)=\bm{a}_{T,i}(\theta_i)=\bm{a}_{i}(\theta_i)$, where the steering vector is expressed as 
  \begin{eqnarray}
 \bm{a}_{i}(\theta_i)=\big[1,e^{j\frac{2\pi}{\lambda}d_i\sin(\theta_i)},\cdots,e^{j\frac{2\pi}{\lambda}d_i(K-1)\sin(\theta_i)}\big]^T,
  \end{eqnarray}
where $d_i$ is the antenna spacing and $\lambda$ is the wavelength. 
 \subsection{Communication and Computation Model}
In the mobile crowd ISCC system, all ISAC devices share the common channel resources. Therefore, when they transmit the ISAC signals to the BS, the received signal at the BS from ISAC device~$i$ is expressed as 
\begin{eqnarray}\label{communication signal}
&&\bm{y}_{c,i}(t)=\bm{H}_i(t)\bm{x}_i(t)+\sum_{j\neq i,j\in\mathcal{N}}\bm{H}_j(t)\bm{x}_j(t)+\bm{v}(t)\nonumber\\
&&=\bm{H}_i(t)\bm{w}_{c,i}(t)s_{c,i}(t)+\underbrace{\sum_{j\neq i,j\in\mathcal{N}}\bm{H}_j(t)\bm{w}_{c,j}(t)s_{c,j}(t)}_{\begin{aligned}
		 &\small\text{Interference from communica-}\\
		&\small\text{tion signals of other ISAC devices}
		\end{aligned}}\nonumber\\
&&~~~+\underbrace{\sum_{j\in\mathcal{N}}\bm{H}_j(t)\bm{w}_{r,j}(t)s_{r,j}(t)}_{\begin{aligned}
		&\small\text{Interference from sensing}\\
		&\small\text{signals of all ISAC devices}
	\end{aligned}}+\bm{v}(t),
\end{eqnarray}
where $\bm{H}_i(t)\in\mathbb{C}^{M\times K}$ is the channel state information~(CSI) between ISAC device $i$ and the BS, and $\bm{v}(t)$ is an additive white Gaussion noise vector with covariance $\sigma_c^2\bm{I}_M$.
%For each sensor, the joint communication signal and radar sensing waveform is utilized for sensing, thus the covariance of the transmission waveform is derived as 
%\begin{eqnarray}
%	&&\bm{R_i}(t)=\mathbb{E}[\bm{x}_i(t)\bm{x}_i^H(t)]\nonumber\\
%	&&~~~~~~~=\bm{w}_{r,i}(t)\bm{w}_{r,i}^H(t)+\bm{w}_{c,i}(t)\bm{w}_{c,i}^H(t).
%\end{eqnarray}
%Therefore, the corresponding power of the waveform used for sensing and communication is expressed as
%\begin{eqnarray}
% P_i(t)=\text{Tr}\big(\bm{R}_i(t)\big).
%\end{eqnarray}
Based on~\eqref{communication signal}, the signal to interference plus noise ratio~(SINR) in the transmission process for ISAC device $i$ at time slot $t$ can be calculated as 
\begin{eqnarray}\label{SNR}
	\chi_i(t)=\frac{\arrowvert \bm{H}_i^H(t)\bm{w}_{c,i}(t)\arrowvert^2}{\arrowvert\bm{n}_i\arrowvert^2+\sigma_c^2},
\end{eqnarray}
where $\bm{n}_i$ is  the interference from other ISAC devices, which is expressed as
\begin{eqnarray}\label{interference}
	\bm{n}_i=\sum_{j\neq i,j\in\mathcal{N}}\bm{H}_j^H(t)\bm{w}_{c,j}(t)+\sum_{j\in\mathcal{N}}\bm{H}_j^H(t)\bm{w}_{r,j}(t).
\end{eqnarray}
Based on the SINR in~\eqref{SNR}, the transmission rate for ISAC device $i$ is derived as 
\begin{eqnarray}\label{data_transmission}
	R_i(t)=B\log_2\big(1+\chi_i(t)\big),
\end{eqnarray}
where $B$ is the transmission bandwidth.

A large volume of data is generated and collected when the ISAC devices perform sensing. The data needs to be processed and analyzed timely. The sensing signal received by ISAC devices is converted into a set of sensing data. Based on the practical engineering practice and radar detection model described in~\cite{ding2022joint,farina2006introduction}, the amount of data size for ISAC devices $i$ is 
\begin{eqnarray}
	D_i(t)=\hat{\zeta}_i\nu_i N_i^{\theta} f_i^s(t)b_i,
\end{eqnarray}
where $\hat{\zeta}_i\geq 1$ is a constant related to the data redundancy, $\nu_i$ is the switching speed of the beam for ISAC device $i$, $N_i^{\theta}$ is the number of quantized angles, $f_i^s(t)$ is the sampling frequency, and $b_i$ is the number of quantization bits for each sample for ISAC device $i$.

The sensing data collected by ISAC devices will be transmitted to the BS for computation, and thus the computational cost should be considered. The computing resources of MEC servers are combined to create a MEC pool via the computation virtualization techniques~\cite{liang2019multiuser}, which can provide sufficient computing resources for ISAC devices. In the MEC pool, each task performed by an ISAC device can be assigned to a virtual machine with a specific CPU processing frequency by using computational virtualization techniques~\cite{zhang2017fog,ku20175g}. Because all ISAC devices share the same physical CPU, the amount of data they offload to the MEC server determines the working frequency of the serving CPU. This results in a quadratic cost for the total load on the CPU. Using the cost model and auction model~\cite{zheng2021dynamic}, the unit computational price for each ISAC device can be defined as 
\begin{eqnarray}\label{computation_cost}
\Phi_i(t)=\left\{\begin{array}{ll}  ~~~~~~~~\kappa,~~~~~~~~~~~~~~~~~~~~~N=1,\\ 
\kappa+\frac{\varrho}{N-1}\sum_{j\in\mathcal{N},j\neq i}R_j(t),~~N\geq2,
\end{array}\right. 
\end{eqnarray}
where $\kappa$ is the basic price for the MEC pool to process per unit data, and $\varrho$ is employed to convert the average load to monetary value.
\begin{remark}
The computational price in (10) reflects the supply and demand relationship of the computing resources. As the data transmitted to servers for computing increases, it leads to the scarcity of computational resources, increasing computing unit price, and an escalation in computational costs.	
\end{remark}
\subsection{Data Dynamics Process}
In this paper, the data is generated by sensing and transmitted by communication for computation. Therefore, the dynamics of data queue for ISAC device $i$ is expressed as 
\begin{eqnarray}\label{data queue dynamics}
	dq_i(t)=\Big(-R_i(t)+D_i(t)\Big)dt.
\end{eqnarray}
The queue dynamic in~\eqref{data queue dynamics} shows the process of data collection and transmission for ISAC devices in sensing and communication processes. At first, each ISAC device collects the data through sensing and then transmits the data to the BS for computation. The two steps constitute the arrival and departure of data in the queue, respectively.
\begin{remark}[Difficult to calculate the influence from other ISAC devices]
As indicated in~\eqref{interference}, the influence from other ISAC devices affects the transmission rate. Therefore, when each ISAC device minimizes its cost by designing the waveform precoding, it should estimate the influence from other ISAC devices. However, in the large-scale mobile crowd ISCC system, each ISAC device cannot obtain the detailed information about other ISAC devices, which makes it difficult to calculate the SINR. Consequently, it is non-trivial to calculate the influence from other ISAC devices in such a large-scale mobile crowd ISCC system.
\end{remark}
\section{Problem Formulation}\label{section_problem_formulation}
In this section, we first formulate the cost function for each ISAC device in the mobile crowd ISCC system in Section~\ref{section_problem_formulation}-A, which consists of the energy consumption and computational cost. Then, we formulate the waveform precoding design problem in Section~\ref{section_problem_formulation}-B.
\subsection{Cost Function for ISAC Devices}
The cost function for each ISAC device is considered from two aspects: energy consumption and computational cost. Energy efficiency is an important factor for the wireless network~\cite{huang2019reconfigurable}, especially when the network resources are scarce. In this paper, we consider the energy consumption in sensing and communication processes, where the energy consumption of ISAC device~$i$ is calculated as  
\begin{eqnarray}\label{energy_consumption}
	J_i^1(\bm{W}_i)=\beta_1\int_{0}^{T}\|\bm{W}_i(t)\|_F^2~dt,
\end{eqnarray}
where $\beta_1$ is the corresponding weight for the energy consumption and $T$ is the total time duration for the ISAC device to finish the sensing task, i.e., the tolerable task delay.

When the tasks are transmitted to the BS, it will consume the computation resources. The computational cost is the fee paid to the server, which is derived as
\begin{eqnarray}\label{computational_cost}
	J_i^2(\bm{W}_i)=\beta_2\int_{0}^{T}\Phi_i(t)R_i(t)dt,
\end{eqnarray}
where $\beta_2$ is the corresponding weight for computational cost. 

Based on the derivation of energy consumption and computational cost, the cost function of ISAC device $i$ is defined as\footnote{We consider a multi-objective optimization problem and utilize the corresponding weights to unify the energy consumption and computational cost functions and obtain a composite objective function in the paper.}
\begin{eqnarray}\label{value_function}
	&&J_i(\bm{W}_i)=J_i^1(\bm{W}_i)+J_i^2(\bm{W}_i)\nonumber\\
	&&~~=\int_{0}^{T}\beta_1\|\bm{W}_i(t)\|_F^2+\beta_2\Phi_i(t)R_i(t)dt.
\end{eqnarray}

\subsection{Waveform Precoding Design Problem for ISAC Devices}
To formulate the waveform precoding design problem, we firstly define $P_{d,i}(\theta_l)$ as the ideal beampattern of ISAC device $i$, where $\theta_l$ is the $l$th sampled angle. We adopt the mean squared error~(MSE) method to evaluate the beampattern similarity~\footnote{The sensing process is also affected by other ISAC devices, as shown in~\eqref{simply_received_radar_signal}. For the ISAC system, the SINR is also a metric to measure the sensing performance~\cite{rihan2018optimum}, which is similar to the communication process. We can extend the proposed problem formulation by adding the sensing SINR as a constraint, and solve the extended problem using the same algorithm by adding a Lagrange multiplier, which will be studied in future work.}, which reflects the sensing performance~\cite{luo2022joint} and is expressed as 
\begin{eqnarray}
&&\text{MSE}(\gamma_i,\bm{W}_i(t))=\frac{1}{L}\sum_{l=1}^L\mid\gamma_iP_{d,i}(\theta_l)\nonumber\\
&&~~~~~~~-\bm{a}_i^H(\theta_l)\bm{W}_i(t)\bm{W}_i^H(t)\bm{a}_i(\theta_l)\mid^2,
\end{eqnarray}
where $\gamma_i$ is a scaling factor for ISAC device $i$.

In this paper, each ISAC device designs the waveform precoding to minimize its energy consumption and computational cost, the corresponding optimization problem is formulated as 
\begin{eqnarray}\label{optimization_problem}
		&&\mathcal{P}:\min_{\bm{W}_i,\gamma_i}\int_{0}^{T}\beta_1\|\bm{W}_i(t)\|_F^2+\beta_2\Phi_i(t)R_i(t)dt+Cq_i(T)\nonumber\\
	&&\text{s.t.}\left\{\begin{array}{ll} 
		\mathcal{C}1:\text{MSE}(\gamma_i,\bm{W}_i(t))\leq \zeta,\\
		\mathcal{C}2:\|\bm{W}_i(t)\|_F^2\leq P_{\max},\\
		\mathcal{C}3:dq_i(t)=\big(-R_i(t)+D_i(t)\big)dt,\\
		\mathcal{C}4:q_i(0)=0,
	\end{array}\right.
\end{eqnarray}
where $\zeta$ denotes the waveform precoding design threshold, which represents the sensing accuracy, $P_{\max}$ is the budget of the total power consumption, and $C$ is a large constant that acts as a penalty factor to ensure that the data state is zero at the terminal time $T$. The problem formulation in~\eqref{optimization_problem} encompasses various optimizations because the constrained optimization problems share the same Lagrangian function~\cite{wang2022distributed}. In problem $\mathcal{P}$, constraint $\mathcal{C}1$ guarantees the sensing performance for each ISAC device, $\mathcal{C}2$ represents the power constraint for each ISAC device, $\mathcal{C}3$ denotes the data queue dynamic, and $\mathcal{C}4$ is the initial condition.

To solve problem $\mathcal{P}$, the unit price for computation $\Phi_i(t)$ and the transmission rate $R_i(t)$  are needed. However, calculating the parameters requires information from all the other ISAC devices, as shown in~\eqref{data_transmission} and~\eqref{computation_cost}. It is non-trivial to formulate the influence from other ISAC devices in such a large-scale mobile crowd ISCC system due to the complex interactions among ISAC devices. Therefore, we utilize the MFG to solve this problem by modeling the interaction of a certain ISAC device with the collective effect of the ISAC devices rather than all the other ISAC devices. The detail of the MFG is shown in the next section.
\section{Problem Reformulation based on Mean Field Game}\label{section_problem_reformualtion}
In this section, we reformulate the waveform precoding design problem based on the MFG, which can reduce the computational complexity of the algorithm significantly. Specifically, we first introduce the concept of MFG in Section~\ref{section_problem_reformualtion}-A, and then reformulate the optimization problem in Section~\ref{section_problem_reformualtion}-B.
\subsection{Introduction to MFG}\label{subsection_introduction_MFG}
In this subsection, we introduce the basic idea of the MFG. The main idea of MFG is that each ISAC device has a negligible effect on the system in the large-scale system, therefore, a certain player reacts to the collective behavior of the players rather than all the other players in the system~\cite{banez2021mean}, which is different from traditional game theory. The collective behavior is reflected in the mean field term, which will be introduced later. Similar to the other game model, the elements of MFG are expressed as
	\begin{itemize}
		\item \emph{Agents}: In our game, the agents are the ISAC devices, denoted as $\mathcal{N}=\{1,2,\cdots,N\}$.
		\item \emph{State}: The state of ISAC devices is the data queue, denoted as $\mathcal{S}(t)=\big\{q_1(t),q_2(t),\cdots,q_N(t)\big\}$.
		\item \emph{Action}: Define $\mathcal{W}(t)=\big\{\bm{W}_1(t),\bm{W}_2(t),\cdots,\bm{W}_N(t)\big\}$ as  the sets of actions, where $\bm{W}_i(t)$ is the waveform precoding design for ISAC device $i$ at time slot $t$.
		\item \emph{Cost function}: As derived in~\eqref{value_function}, the cost function $J_i(\bm{W}_i)$ is combined with the energy consumption and the computation cost. 
	\end{itemize}	
	Since the number of ISAC devices is huge, the state of the data queue can be approximated by its probability distribution, known as the mean field term~(MFT)~\cite{wang2021delay}, which is defined as 
	\begin{eqnarray}
		\rho(q,t)=\frac{1}{N}\sum_{i\in\mathcal{N}}\mathbf{1}_{\{q_i(t)=q\}},
	\end{eqnarray}
	where $\mathbf{1}_{\{q_i(t)=q\}}=1$ if $q_i(t)=q$, and $0$ otherwise.
	
MFT is a crucial concept in MFG, playing a significant role in simplifying the interactions among ISAC devices. It involves utilizing a distribution over the state space of ISAC devices to represent the collective state of all ISAC devices. To derive the optimal strategy $\bm{W}_i(t)$, the following partial differential equations~(PDEs) should be solved~\cite{lasry2007mean,kang2020joint}:
\begin{eqnarray}
	&&-\frac{\partial V}{\partial t}+\hat{H}\bigg(q,\rho,\frac{\partial V}{\partial\bm{W}}\bigg)=0,\label{HJB_equation}\\
	&&\partial_t\rho(q,t)+\partial_q(\rho(q,t)\partial_tq)=0,\label{FPK_equation}\\
	&&\rho(q,0)=\rho_0,V(q,T)=V_T,
\end{eqnarray}
where $\hat{H}\left(q,\rho,\frac{\partial V}{\partial\bm{W}}\right)$ is the Hamilton equation and $V(q,t)$ is the value function, which is defined as
\begin{eqnarray}
	V(q,t)=\min_{\bm{W}(t)}J(q,\bm{W}).
\end{eqnarray}
Equation \eqref{HJB_equation} is the Hamiltion-Jacobi-Bellman~(HJB) equation, and \eqref{FPK_equation} is the FPK equation, where the HJB equation is widely known, thus its derivation is omitted here. The derivation of the FPK equation is shown in Appendix A. By solving the aforementioned PDEs, we can obtain the optimal waveform precoding $\bm{W}$.

\subsection{MFG Problem Reformulation}
In this part, we will reformulate the optimization problem into an MFG problem. We first derive the cost function and transform the constraints. Then, the reformulated problem is introduced.

\subsubsection{Cost function reformulation}
Based on MFG, we can reformulate the average energy consumption for each ISAC device as
\begin{eqnarray}\label{reformulation_energy_cost}
	&&\bar{J}^1(q,\bm{W})=\mathbb{E}\bigg[\beta_1\int_{0}^{T}\|\bm{W}(q,t)\|_F^2dt\bigg]\nonumber\\
	&&~~~~~~~~~~=\beta_1\int_{0}^{T}\int_{\mathcal{S}}\|\bm{W}(q,t)\|_F^2\rho(s,t)~dqdt,
\end{eqnarray}
where $\mathcal{S}$ is the state space of the data queue. As shown in~\eqref{reformulation_energy_cost}, the energy consumption for each ISAC device is the integral of the energy consumption over $[0,T]$ and state space $\mathcal{S}$.

On the other hand, the average computational cost for each ISAC device also should be reformulated. In this large-scale mobile crowd ISCC system, i.e., $N\gg2$, the computational cost in~\eqref{computation_cost} can be reformulated into
\begin{eqnarray}\label{computation_cost_reformulation}
&&\Phi(t)=\kappa+\varrho\bigg[\frac{N}{N-1}\int_{\mathcal{S}}R(q,t)\rho(q,t)dq\nonumber\\
&&~~~~~~~~-\frac{1}{N-1}R(q,t)\bigg].
\end{eqnarray}
In~\eqref{computation_cost_reformulation}, when $N\rightarrow\infty$, $\frac{N}{N-1}\rightarrow1$ and $\lim_{N\rightarrow\infty}\frac{1}{N-1}R(q,t)=0$ since $R(q,t)$ is bounded. Therefore, the computational cost $\Phi(q,t)$ can be approximated by
\begin{eqnarray}
	\Phi(t)=\kappa+\varrho\int_{\mathcal{S}}R(q,t)\rho(q,t)dq.
\end{eqnarray}
Then, we calculate the influence from other ISAC devices in~\eqref{SNR}, which can be derived as 
\begin{eqnarray}
&&\arrowvert\bm{n}(t)\arrowvert^2=N\int_{\mathcal{S}}\arrowvert\bm{H}^H(t)\bm{w}_{c}(q,t)\arrowvert^2\rho(q,t)dq\nonumber\\
&&~~~~+N\int_{\mathcal{S}}\arrowvert\bm{H}^H(t)\bm{w}_{r}(q,t)\arrowvert^2\rho(q,t)dq,
\end{eqnarray}
where $\bm{w}_r(q,t)$ and $\bm{w}_c(q,t)$ are the waveform precoding design of sensing signal and communication signal, respectively, $\bm{H}(t)$ is the corresponding CSI. Here, we utilize the state to identify ISAC devices and omit the subscript of ISAC devices in the following part.

Then, the average computational cost is derived as 
\begin{eqnarray}\label{reformulated_computation_cost}
&&\bar{J}^2(q,\bm{W})=\mathbb{E}\bigg[\int_0^T\Phi_i(t)R_i(t)dt\bigg]\nonumber\\
&&~~~~~~~=\int_0^T\int_{\mathcal{S}}\beta_2\Phi(t)R(q,t)\rho(q,t) dqdt.
\end{eqnarray}
Given $\bar{J}^1(q,\bm{W})$ and $\bar{J}^2(q,\bm{W})$, the cost function is reformulated into 
\begin{eqnarray}
	&&\bar{J}(q,\bm{W})=\bar{J}^1(q,\bm{W})+\bar{J}^2(q,\bm{W})\nonumber\\
	&&~~=\int_0^T\int_{\mathcal{S}}\Big[\beta_1\|\bm{W}(q,t)\|_F^2+\beta_2\Phi(q,t)R(q,t)\Big]\nonumber\\
	&&~~~~~~\cdot\rho(q,t) dqdt+\int_{\mathcal{S}} Cq(T)\rho(q,T)dq.
\end{eqnarray}
\subsubsection{Constraint transformation} Next, we will transform the constraints for utilizing the G-prox PDHG algorithm in the next section. 

At first, we transform the sensing performance constraint in~$\mathcal{C}1$. Since the variable $\gamma$ only exists in constraint $\tilde{\mathcal{C}}1$, and left hand side~(l.h.s.) is a quadratic and convex function with respect to (w.r.t.) $\gamma$. Therefore, we can obtain $\gamma^*$ by calculating $\frac{\partial \text{MSE}(\gamma,\bm{W}(q,t))}{\partial \gamma}$, and then we have
\begin{eqnarray}
	\gamma^*=\frac{\sum_{l=1}^LP_{d}(q,\theta_l)\text{vec}^H\big(\bm{A}_l\big)\text{vec}\big(\bm{W}(q,t)\bm{W}^H(q,t)\big)}{\sum_{l=1}^LP_{d}^2(q,\theta_l)},
\end{eqnarray}
where $\bm{A}_l=\bm{a}(\theta_l)\bm{a}^H(\theta_l)$.
We substitute $\gamma^*$ into $\text{MSE}(\gamma,\bm{W}(q,t))$, and then constraint $\mathcal{C}1$ can be re-expressed as 
\begin{eqnarray}\label{transform MSE}
&&\text{MSE}({\bm{W}(q,t)})=\text{vec}^H(\bm{W}(q,t)\bm{W}^H(q,t))\bm{C}\nonumber\\
&&~~~~~\times\text{vec}(\bm{W}(q,t)\bm{W}^H(q,t))\leq\zeta,
\end{eqnarray}
where
\begin{eqnarray}
	&&\bm{C}=\frac{1}{L}\sum_{l=1}^L\bm{b}_l\bm{b}_l^H,\label{definition_C}\\
	&& \bm{b}_l=\frac{P_d(\theta_l)\sum_{l_1=1}^LP_d(\theta_{l_1})\text{vec}(\bm{A}_{l_1})}{\sum_{l_1=1}^LP_d^2(\theta_{l_1})}-\text{vec}(\bm{A}_{l})\label{definition_b}.
\end{eqnarray}
Since the constraint in~\eqref{transform MSE} is quartic and non-convex, we transform $	\text{MSE}({\bm{W}(q,t)})$ into a series of functions by utilizing MM method~\cite{luo2022joint}. Given the value $\bm{W}^k(q,t)$ at the $k$th update, we construct a more tractable proxy function to approximate $\text{MSE}(\bm{W}(q,t))$ and serve as its upper bound, which is expressed as 
\begin{eqnarray}\label{MSE_inequality}
	&&\text{MSE}({\bm{W}(q,t)})\leq\lambda_m\text{vec}^H(\bm{W}(q,t)\bm{W}^H(q,t))\nonumber\\
   &&~~~~\times\text{vec}(\bm{W}(q,t)\bm{W}^H(q,t))+\mathbb{R}\{\text{vec}(\bm{W}(q,t)\nonumber\\
	&&~~~~\times\bm{W}^H(q,t))\bm{b}^k\}+e_1^k,
\end{eqnarray}
where 
\begin{eqnarray}
	&&\bm{b}^k=2(\bm{C}-\lambda_m\bm{I})\text{vec}(\bm{W}^k(q,t)(\bm{W}^k(q,t))^H),\label{definition_bt}\\
	&&e_1^k=\text{vec}^H(\bm{W}^k(q,t)(\bm{W}^k(q,t))^H)(\lambda_m\bm{I}-\bm{C})\nonumber\\
&&~~~~\times\text{vec}(\bm{W}^k(q,t)(\bm{W}^k(q,t))^H),
\end{eqnarray}
where $\lambda_m$ is the maximum eigenvalue of $\bm{C}$. Based on the power constraint in $\mathcal{C}_2$, the upper bound of the first term on the right-hand side~(r.h.s.) of~\eqref{MSE_inequality} is derived as 
\begin{eqnarray}\label{inequality2}
	&&\lambda_m\text{vec}^H(\bm{W}(q,t)(\bm{W}(q,t))^H)\text{vec}(\bm{W}(q,t)(\bm{W}(q,t))^H)\nonumber\\
	&&=\lambda_m\bigg\|\sum_{j=1}^2\bm{w}_j\bm{w}_j^H\bigg\|_F^2=\lambda_mP_{\max}^2.
\end{eqnarray}                  
where $\bm{w}_j$ is the $j-$th column of $\bm{W}$. By substituting~\eqref{inequality2} into~\eqref{MSE_inequality}, we can obtain the upper bound of $\text{MSE}({\bm{W}(q,t)})$, which is expressed 
\begin{eqnarray}\label{inequality3}
&&\text{MSE}({\bm{W}(q,t)})\leq\mathbb{R}\{\text{vec}(\bm{W}(q,t)\bm{W}^H(q,t))\bm{b}^k\}\nonumber\\
&&~~~~~+e_1^k+\lambda_mP_{\max}^2.
\end{eqnarray} 
Further, we rewrite $\mathbb{R}\{\text{vec}(\bm{W}(q,t)\bm{W}^H(q,t))\bm{b}^k\}$ as
\begin{eqnarray}\label{derive 1}
	&&\mathbb{R}\{\text{vec}(\bm{W}(q,t)\bm{W}^H(q,t))\bm{b}^k\}=\sum_{j=1}^2\mathbb{R}\{\text{vec}^H(\bm{w}_j\bm{w}_j^H)\bm{b}^k\}\nonumber\\
	&&~~~~~~~~~~~~=\sum_{j=1}^2\mathbb{R}\{\bm{w}_j^H\bm{B}^k\bm{w}_j\},
\end{eqnarray}
where $\bm{B}^k\in\mathbb{C}^{N\times N}$ is a reshaped version of $\bm{b}^t$ and satisfies $\bm{b}^k=\text{vec}(\bm{B}^k)$. Based on~\eqref{definition_C}, \eqref{definition_b} and~\eqref{definition_bt}, $\bm{B}^k$ can be divided into 
\begin{eqnarray}
	\bm{B}^k=\bm{B}_1^k+\bm{B}_2^k,
\end{eqnarray}
where 
\begin{eqnarray}
	&&\bm{B}_1^k=\frac{2}{L}\sum_{l_1=1}^L\frac{P_d^2(\theta_{l_1})}{\Big(\sum_{l=1}^LP_d^2(\theta_l)\Big)^2}\sum_{l_2=1}^LP_d^2(\theta_{l_2})\text{vec}^H(\bm{A}_{l_2})\nonumber\\
&&~~~~\times\text{vec}(\bm{W}(q,t)\bm{W}^H(q,t))\sum_{l_3=1}^L P_d^2(\theta_{l_3})\bm{A}_{l_3}\nonumber\\
	&&~~~~+\frac{2}{L}\sum_{l_1=1}^L\text{vec}^H(\bm{A}_{l_1})\text{vec}(\bm{W}(q,t)\bm{W}^H(q,t))\bm{A}_{l_1},\label{derive 2}\\
	&&\bm{B}_2^k=-\frac{4}{L}\mathbb{R}\Bigg\{\sum_{l_1=1}^L\frac{P_d(\theta_{l_1})}{\sum_{{l}=1}^LP_d^2(\theta_l)}\text{vec}^H(\bm{A}_{l_1})\nonumber\\
&&~~~~\times\text{vec}(\bm{W}(q,t)\bm{W}^H(q,t))\sum_{{l_2}=1}^LP_d(\theta_{l_2})\bm{A}_{l_2}\Bigg\}\nonumber\\
	&&~~~~-2\lambda_m\bm{W}^k(q,t)(\bm{W}^k(q,t))^H.\label{derive 3} 
\end{eqnarray}
Then, the function $\bm{w}_j^H\bm{B}^k\bm{w}_j$ can be divided into
\begin{eqnarray}\label{shaped_version}
	\bm{w}_j^H\bm{B}^k\bm{w}_j=\bm{w}_j^H\bm{B}_1^k\bm{w}_j+\bm{w}_j^H\bm{B}_2^k\bm{w}_j.
\end{eqnarray} 
Since $\bm{B}_1^k$ is a positive semidefinite matrix and $\bm{B}_2^k$ is a negative semidefinite matrix,  the first term on the r.h.s. of~\eqref{shaped_version} is a convex function, while the second term is a concave function. Therefore, we also utilize the MM method to find a convex function for the concave function in each iteration. When the first Taylor expansion is utilized, a convex upper bound of the concave function $\bm{w}_j^H\bm{B}_2^k\bm{w}_j$ is expressed as 
\begin{eqnarray}\label{derive 4} 
	&&\bm{w}_j^H\bm{B}_2^k\bm{w}_j\leq(\bm{w}_j^k)^H\bm{B}_2^k\bm{w}_j^k\nonumber\\
&&~~~~+2\mathbb{R}\big\{(\bm{w}_j^k)^H\bm{B}_2^k\bm{w}_j^k(\bm{w}_j-\bm{w}_j^k)\big\},
\end{eqnarray}
where $\bm{w}_j^k$ is the $j$th column of $\bm{W}^k$ at the $k$th iteration.
Substituting~\eqref{derive 1}-\eqref{derive 4} into~\eqref{inequality3}, a convex upper bound of $\text{MSE}({\bm{W}(q,t)})$ is derived as
\begin{eqnarray}
	\text{MSE}({\bm{W}(q,t)})\leq\sum_{j=1}^2\mathbb{R}\{\bm{w}_j^H\bm{B}_1^k\bm{w}_j+2\bm{w}_j^H\bm{u}_j\}+e_2^k,
\end{eqnarray}
where 
\begin{eqnarray}
	&&\bm{u}_j=(\bm{B}_2^k)^H\bm{w}_j^k,\\
	&&e_2^k=-\mathbb{R}\{(\bm{w}_j^k)^H(\bm{B}_2^k)^H\bm{w}_j^k\}+e_1^k+\lambda_mP_{\max}^2.
\end{eqnarray}

Then, we transform the queue dynamic constraint in $\mathcal{C}3$. As we introduced in Section~\ref{subsection_introduction_MFG}, the FPK equation can model the evolution of the system state, therefore, the queue dynamic constraints can be transformed into a signal FPK equation in~\eqref{FPK_equation}. However, the FPK equation is not a linear constraint due to the logarithm function in~\eqref{data_transmission}, we also need to handle it. The SINR can approximate to a specific value, denoted as $x_0$, when we adopt a coding method~\cite{schlegel2015trellis}. Therefore, we utilize the first order Taylor expansion to handle $\partial_tq$, i.e.,
\begin{eqnarray}\label{taylor_expansion}
	&&R(q,t)=B\Big(\log_2(1+x_0)+\frac{1}{\ln2(1+x_0)}(\chi(q,t)\nonumber\\
&&~~~~-x_0)\Big)+o(\chi(q,t)-x_0)^2.
\end{eqnarray}
Based on~\eqref{taylor_expansion}, the FPK equation is transformed into
\begin{eqnarray}
	&&\partial_t\rho(q,t)+\partial_q(\rho(q,t)\Gamma(q,t))=0,
\end{eqnarray}
where $\Gamma(q,t)=-B(\log_2(1+x_0)+1/(\ln2(1+x_0))(\chi(q,t)-x_0))+D(t)$.

Consequently, the reformulated MFG problem $\mathcal{P}1$ is given by 
\begin{eqnarray}
	&&\mathcal{P}1:\min_{\rho,\bm{W}}~~~~~\bar{J}(q,\bm{W})\nonumber\\
	&&\text{s.t.}\left\{\begin{array}{ll} 
		\tilde{\mathcal{C}}1:\sum_{j=1}^{2}\mathbb{R}\{\bm{w}_j^H\bm{B}_1^k\bm{w}_j+2\bm{w}_j^H\bm{u}_j\}+e_2^k\leq \zeta,\\
		\tilde{\mathcal{C}}2:\|\bm{W}(q,t)\|_F^2\leq P_{\max},\\
		\tilde{\mathcal{C}}3:\partial_t\rho(s,t)+\partial_q(\rho(q,t)\Gamma(q,t))=0.
	\end{array}\right.
\end{eqnarray}

In problem $\mathcal{P}1$, constraints $\mathcal{C}1$ and $\mathcal{C}2$ in~\eqref{optimization_problem} for each ISAC device are transformed into $\tilde{\mathcal{C}}1$ and $\tilde{\mathcal{C}}2$ for each state~$q$. The main difference is reflected in the transformation of constraint $\mathcal{C}3$. Since the FPK equation models the evolution of system state, which is exactly the data queue dynamic in constraint $\mathcal{C}3$. Consequently, we utilize the FPK equation to replace the queue dynamic and the initial condition is also correspondingly changed into $\rho(q,0)=\rho_0$. We will design the waveform precoding algorithm to solve the reformulated problem $\mathcal{P}1$ in the next section.
\section{Waveform Precoding Design Algorithm}
In this section, we utilize the G-prox PDHG algorithm to solve the reformulated waveform precoding design problem. Specifically, we first give the details of the G-prox PDHG algorithm in Section~\ref{subsection_detail_PDHG}. Then, the convergence, uniqueness, as well as computational complexity of the proposed waveform precoding design algorithm are analyzed in Section~\ref{subsection_complexity analysis}.

\subsection{Details of the G-prox PDHG Algorithm}\label{subsection_detail_PDHG}
In this subsection, we give the solution to problem $\mathcal{P}1$ based on the G-prox PDHG algorithm. To show the details of the G-prox PDHG algorithm, we first give the Lagrangian function $L(\bm{W},\rho,\phi_1,\phi_2,\phi_3)$ of problem $\mathcal{P}1$ as follows:
\begin{eqnarray}\label{Lagrangian_function}
&&L(\bm{W},\rho,\phi_1,\phi_2,\phi_3)\nonumber\\
&&=\bar{J}(q,\bm{W})+\int_0^T\int_{\mathcal{S}}\phi_1(q,t)c_1(q,t)+\phi_2(q,t)c_2(q,t)\nonumber\\
&&~~~+\phi_3(q,t)(\partial_t\rho(q,t)+\partial_q\cdot(\rho(q,t)\Gamma(q,t))dqdt,
\end{eqnarray}
where $\phi_1(q,t)$, $\phi_2(q,t)$, and $\phi_3(q,t)$ are the Lagrangian dual variables corresponding to constraints $\tilde{\mathcal{C}}1$, $\tilde{\mathcal{C}}2$, and $\tilde{\mathcal{C}}3$, respectively, $c_1\overset{\triangle}{=}\sum_{j=1}^{2}\mathbb{R}\{\bm{w}_j^H\bm{B}_1^k\bm{w}_j+2\bm{w}_j^H\bm{u}_j\}+e_2^k-\zeta$, and  $c_2\overset{\triangle}{=}\|\bm{W}(q,t)\|_F^2-P_{\max}$.  Based on the defined Lagrangian function, we are committed to solving the following min-max problem 
\begin{eqnarray}\label{min-max_problem}
\min_{\rho,\bm{W}}\Big\{\max_{\phi_1,\phi_2,\phi_3}L(\bm{W},\rho,\phi_1,\phi_2,\phi_3)\Big\}.
\end{eqnarray}
To solve the above problem, we first initialize $\rho(q,t)$, $\bm{W}(q,t)$, $\phi_1(q,t)$, $\phi_2(q,t)$, and $\phi_3(q,t)$ for $q\in\mathcal{S}$ and $t\in[0,T]$, where we define
\begin{eqnarray}\label{initialize-rho0}
	\rho(q,0)=\rho_0.
\end{eqnarray}
Further, the G-prox PDHG algorithm is utilized to solve the min-max problem, and $\rho$, $\bm{W}$, $\phi_1(q,t)$, $\phi_2(q,t)$, and $\phi_3(q,t)$ can be obtained by
\begin{eqnarray}
&&\rho^{k+1}=\arg\min_{\rho}\mathcal{L}_{\rho}\nonumber\\
&&~~~~~~=\arg\min_{\rho}\Big\{L(\bm{W}^k,\rho,\phi_1^k,\phi_2^k,\phi_3^k)\Big\}\nonumber\\
&&~~~~~~~~~~+\frac{1}{2\xi}\|\rho-\rho^k\|_{L^2}^2,\\
&&\bm{W}^{k+1}=\arg\min_{\bm{W}}\mathcal{L}_{\bm{W}}\nonumber\\
&&~~~~~~~=\arg\min_{\rho}\Big\{L(\bm{W},\rho^{k+1},\phi_1^k,\phi_2^k,\phi_3^k)\Big\}\nonumber\\
&&~~~~~~~~~~+\frac{1}{2\xi}\|\bm{W}-\bm{W}^k\|_{L^2}^2,\\
&&\phi_1^{k+1}=\arg\min_{\phi_1}\mathcal{L}_{\phi_1}\nonumber\\
&&~~~~~~=\arg\min_{\phi_1}\Big\{L(\bar{\bm{W}}^{k+1},\bar{\rho}^{k+1},\phi_1,\phi_2^k,\phi_3^k)\Big\}\nonumber\\
&&~~~~~~~~~~+\frac{1}{2\varsigma}\|\phi_1-\phi_1^k\|_{L^2}^2,\\
&&\phi_2^{k+1}=\arg\min_{\phi_2}\mathcal{L}_{\phi_2}\nonumber\\
&&~~~~~~=\arg\min_{\phi_2}\Big\{L(\bar{\bm{W}}^{k+1},\bar{\rho}^{k+1},\phi_1^{k+1},\phi_2,\phi_3^k)\Big\}\nonumber\\
&&~~~~~~~~~~+\frac{1}{2\varsigma}\|\phi_2-\phi_2^k\|_{L^2}^2,\\
&&\phi_3^{k+1}=\arg\min_{\phi_3}\mathcal{L}_{\phi_3}\nonumber\\
&&~~~~~~=\arg\min_{\phi_3}\Big\{L(\bar{\bm{W}}^{k+1},\bar{\rho}^{k+1},\phi_1^{k+1},\phi_2^{k+1},\phi_3)\Big\}\nonumber\\
&&~~~~~~~~~~+\frac{1}{2\varsigma}\|\phi_3-\phi_3^k\|_{H_1}^2,
\end{eqnarray}
where $\mathcal{L}_{\bm{W}}$, $\mathcal{L}_{\rho}$, $\mathcal{L}_{\phi_1}$, $\mathcal{L}_{\phi_2}$, and $\mathcal{L}_{\phi_3}$ are the augmented Lagrangian function w.r.t. $\rho$, $\bm{W}$, $\phi_1$, $\phi_2$, and $\phi_3$, respectively, $\xi$ and $\varsigma$ are the corresponding step sizes, respectively, $\bar{\bm{W}}^{k+1}=2\bm{W}^{k+1}-\bm{W}^k$, and $\bar{\rho}^{k+1}=2\rho^{k+1}-\rho^k$. The $L_2$ norm is 
\begin{eqnarray}
	\|u(q,t)\|_{L^2}^2=\int_0^T\int_{\mathcal{S}}\arrowvert u(q,t)\arrowvert^2dqdt,
\end{eqnarray}
and the $H_1$ norm is denoted as 
\begin{eqnarray}	\|u(q,t)\|_{H_1}^2=\int_0^T\int_{\mathcal{S}}(\arrowvert \partial_tu(q,t)\arrowvert^2+\arrowvert\partial_q u(q,t)\arrowvert^2)dqdt.
\end{eqnarray}
To update the above variables, we first calculate the corresponding gradients of the Lagrangian function. The gradient w.r.t. $\rho$ and $\bm{W}$ are derived as 
\begin{eqnarray}\label{gradient_of_rho}
&&\frac{\partial \mathcal{L}_{\rho}}{\partial\rho}=\beta_1\|\bm{W}(q,t)\|_F^2+\beta_2\Phi(q,t)R(q,t)+\partial_t\phi_3(q,t)\nonumber\\
&&+\Gamma(q,t)\partial_q(\phi_3(q,t))+\frac{1}{\xi}(\rho(q,t)-\rho^k(q,t)).
\end{eqnarray}
\begin{eqnarray}\label{gradient_of_W}
&&\frac{\partial\mathcal{L}_{\bm{W}}}{\partial\bm{W}}=\nonumber\\
&&\bigg(2\beta_1\bm{W}(q,t)+2\beta_2\frac{B\Phi(q,t)\bm{H}^H(t)\bm{H}(t)\bm{\Xi}(q,t)}{\ln2\arrowvert\bm{H}^H(t)\bm{\Xi}(q,t)\arrowvert^2}\bigg)\nonumber\\
&&\times\rho(q,t)+\frac{1}{2}\phi_1(q,t)(\bm{B}_1^k+(\bm{B}_1^k)^H+2\bm{B}_2^k+2(\bm{B}_2^k)^H)\nonumber\\
&&\times\bm{W}(q,t)+2\phi_2(q,t)\bm{W}(q,t)-\frac{2B\rho(q,t)}{\ln2(1+x_0)}\partial_q\phi_3(q,t)\nonumber\\
&&\times\frac{\bm{H}^H(t)\bm{H}(t)\bm{\Xi}(q,t)}{(\arrowvert\bm{n}\arrowvert^2+\sigma_c^2)}+\frac{1}{\tau}(\bm{W}(q,t)-\bm{W}^k(q,t)),
\end{eqnarray}
where $\bm{\Xi}(q,t)\overset{\triangle}{=}[\bm{0}_{M\times1}~\bm{w}_c(q,t)]$. The related proof can be seen in Appendix B.

Based on the gradient w.r.t. $\rho(s,t)$ in~\eqref{gradient_of_rho}, we update~$\rho$ according to the following equation
\begin{eqnarray}\label{update_rho}
&&\rho^{k+1}(q,t)=\arg\min_{\rho(q,t)}\mathcal{L}_\rho\nonumber\\
&&=\rho^k(q,t)-\xi(\beta_1\|\bm{W}(q,t)\|_F^2+\beta_2\Phi(q,t)R(q,t)\nonumber\\
&&~~~+\partial_t\phi_3(q,t)+\Gamma(q,t)\partial_q(\phi_3(q,t))).
\end{eqnarray}

%\begin{eqnarray}
%	&&\bm{w}_r^{k+1}(q,t)=\Big(\tau\phi_1(q,t)(\bm{B}_1^k+(\bm{B}_1^k)^H+2\bm{B}_2^k+2(\bm{B}_2^k)^H)\nonumber\\
%	&&~~~~+(2\beta_1\rho(q,t)\tau+2\phi_2\tau+1)\bm{I}_K\Big)^{-1}\bm{w}_r^k(q,t),\\
%	&&\bm{w}_c^{k+1}(q,t)=\Big(\tau\phi_1(q,t)(\bm{B}_1^k+(\bm{B}_1^k)^H+2\bm{B}_2^k+2(\bm{B}_2^k)^H)\nonumber\\
%	&&~~~~-\frac{2B\rho(q,t)\bm{H}^H(t)\bm{H}(t)(\beta_2\Phi(q,t)+\triangledown\phi_3(q,t))}{\ln2(1+x_0)(\arrowvert\bm{n}\arrowvert^2+\sigma_c^2)}\nonumber\\
%	&&~~~~+(2\beta_1\rho(q,t)\tau+2\phi_2\tau+1)\bm{I}_K\Big)^{-1}\bm{w}_c^k(q,t)
%\end{eqnarray}
Then, we update $\bm{W}$ by solve~\eqref{gradient_of_W}.
Similarly, we update $\phi_1$, $\phi_2$, and  $\phi_3$ iteratively, which is derived as follows:
\begin{eqnarray}
&&\phi_1^{k+1}=\phi_1^k+\varsigma c_1^k,\label{update_phi_1}\\
&&\phi_2^{k+1}=\phi_2^k+\varsigma c_2^k,\label{update_phi_2}\\
&&\phi_3^{k+1}=\phi_3^k+\varsigma(\triangle)^{-1}(\partial_t\bar{\rho}^{k+1}(q,t)\nonumber\\
&&~~~~~~~+\partial_q(\bar{\rho}^{k+1}(q,t)\Gamma(q,t)),\label{update_phi_3}
\end{eqnarray}
where $c_1^k\overset{\triangle}{=}\sum_{j=1}^{2}\mathbb{R}\{(\bm{w}_j^k)^H\bm{B}_1^k\bm{w}_j^k+2(\bm{w}_j^k)^H\bm{u}_j^k\}+e_2^k-\zeta$, $c_2^k\overset{\triangle}{=}\|\bm{W}^k(q,t)\|_F^2-P_{\max}$, $\bm{u}_j^k(q,t)=(\bm{B}_2^k)^H\bm{w}_j^k(q,t)$, and $(\triangle)^{-1}$ is the inverse Laplace transform.

Note that the $\rho$ update mentioned above does not include the terminal distribution of the state, i.e., $\rho(q,T)$, which should be updated since it is the boundary condition. Similarly, we define the augmented Lagrangian function for $\rho(q,T)$ as
\begin{eqnarray}
&&\rho(q,T)=\arg\min_{\rho(q,T)}\mathcal{L}_{\rho(q,T)}\nonumber\\
&&=\arg\min_{\rho(q,T)}\Big\{L(\bar{\bm{W}}^{k+1},\bar{\rho}^{k+1},\phi_1^{k+1},\phi_2^{k+1},\phi_3^{k+1})\nonumber\\
&&~~~+\frac{1}{2\xi}\|\rho(q,T)-\rho^k(q,T)\|_{L^2}^2\Big\}.
\end{eqnarray}
Therefore, based on the first-order derivative condition, we update $\rho(q,T)$ as follows
\begin{eqnarray}\label{update_rho_T}
\rho^{k+1}(q,T)=\rho^k(q,T)+\xi\big( \phi_3(T)-Cq\big).
\end{eqnarray}

Based on the above update strategies, we utilize the numerical method to obtain the solution. In the proposed numerical algorithm, we discretize the state space $\mathcal{S}$ into $N_1$ grids, and the time interval $[0,T]$ is discretized into $N_2$ grids. Therefore, the corresponding variables are discretized into $\rho(i,j)=\rho(i\triangle q, j\triangle t)$, $\bm{W}(i,j)=\bm{W}(i\triangle q,j\triangle t)$, $\phi_1(i,j)=\phi_1(i\triangle q,j\triangle t)$, $\phi_2(i,j)=\phi_2(i\triangle q,j\triangle t)$, $\phi_3(i,j)=\phi_3(i\triangle q,j\triangle t)$, $\rho(i,T)=\rho(i\triangle q)$.
for $i=1,2,\cdots, N_1$, $j=1,2,~\cdots,~N_2$, and 
\begin{eqnarray}
\triangle q=\frac{Q_{\max}}{N_1},  \triangle t=\frac{T}{N_2}.
\end{eqnarray}
\begin{algorithm}[htbp]
	\caption{G-prox PDHG Algorithm for the Waveform Precoding Design Problem}\label{algorithm_PDHG}
	\begin{algorithmic}[1]
	\STATE  \textbf{Initialize}: $\rho(i,j)$, $\bm{W}(i,j)$, $\phi_1(i,j)$, $\phi_2(i,j)$, and $\phi_3(i,j)$ for $i=1,2,\cdots,N_1$ and $j=1,2,\cdots,N_2$.
		\WHILE{$k\leq K_{\max}$}   
		\FOR{$j=1,2,\cdots,N_2$} 
		\FOR{$i=1,2,\cdots,N_1$} 
		\STATE update $\rho^{k+1}(i,j)$ using~\eqref{update_rho};
		\STATE solve the function in~\eqref{gradient_of_W} and update~$\bm{W}^{k+1}(i,j)$;
		\STATE update $\rho(i,N_2)$ using~\eqref{update_rho_T};
		\STATE update $\phi_1(i,j)$ and $\phi_2(i,j)$ using~\eqref{update_phi_1} and~\eqref{update_phi_2}, respectively;
		\STATE calculate the inverse Laplacian, and update $\phi_3(i,j)$ using~\eqref{update_phi_3};
		\ENDFOR 
		\ENDFOR
		\ENDWHILE 
		\RETURN Optimal waveform precoding design $\bm{W}^*$
	\end{algorithmic}
\end{algorithm}
\subsection{Performance Analysis}
In this part, we analyze the performance of the proposed G-Prox algorithm, including the computational complexity analysis and optimality analysis.
\subsubsection{Computational complexity analysis}\label{subsection_complexity analysis}
In the min-max problem in~\eqref{min-max_problem}, the mean field term $\rho^{k+1}(i,j)$ and the waveform precoding $\bm{W}^{k+1}(i,j)$ are obtained by maximizing the augmented Lagrangian function. As shown in Algorithm~\ref{algorithm_PDHG}, the variables $\rho^{k+1}(i,j)$, $\bm{W}^{k+1}(i,j)$, $\rho^{k+1}(i,T)$, $\phi_1^{k+1}(i,j)$, $\phi_2^{k+1}(i,j)$, $\phi_2^{k+1}(i,j)$ are updated in each iteration. In the discrete solution, the space size is discretized into $N_1\times N_2$ grids. Consequently, we need to update $N_1\times N_2$ times for the variables except for $\rho^{k+1}(i,N_2)$, which is updated only $N_1$ times in each iteration. In addition, the external loop is executed $K$ times. Therefore, the computational complexity is $\mathcal{O}(K\times N_1\times N_2)$, which is linear with the number of space sizes and independent of the number of ISAC devices in the waveform precoding design problem.

\subsubsection{Optimality analysis} 
\begin{figure}[htbp]
{\centering \includegraphics[width=3.1in]{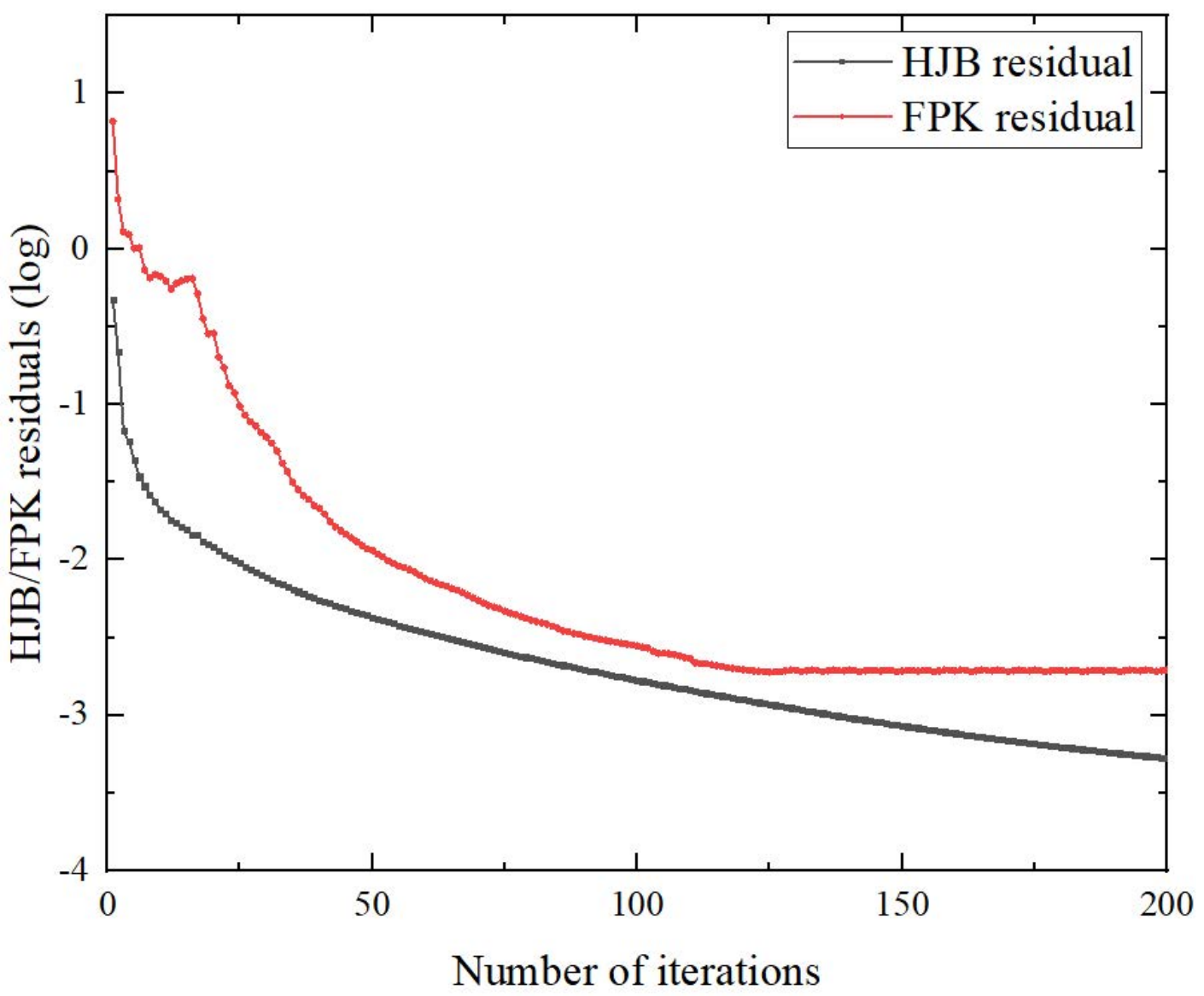}
	\caption{FPK/HJB residuals with the process of iteration}
	\label{fig:residual1}
}
\end{figure}

We develop a numerical method to solve the waveform precoding algorithm, which involves a Lagrangian function and a saddle-point problem. The solution of the numerical method is also the solution to the coupled HJB equation and FPK equation, which is an optimal solution if it satisfies the two coupled PDEs~~\cite{lasry2007mean,kang2020joint}. In addition, to verify the optimality of the proposed algorithm, we have given the simulation result of the residuals of the HJB and FPK equations\footnote{The related parameters can refer to the Section VI.} in Fig.~\ref{fig:residual1}. It can be seen that the solution of the numerical results satisfies the coupled PDEs, which indicates the optimality of the solution.

\section{Simulation Results}
In this section, the simulation results are provided to evaluate the performance of the proposed waveform precoding design algorithm for the large-scale mobile crowd ISCC system. To demonstrate the advantages of the proposed algorithm, we first compare the proposed waveform precoding design algorithm with the existing algorithms in~\ref{subsection_performance_comparison} to demonstrate the superiority of the proposed algorithm. Then, we conduct a comprehensive analysis for the performance of the proposed algorithm in~\ref{subsection_performance_analysis}.
The simulation results are carried out using Matlab 2022b on a server with an AMD EPYC 75F3 32-Core Processor. 

In the simulation, we set the number of ISAC devices as 500 to ensure the large-scale system and assume the number of antennas for ISAC devices and the BS are $K=$2 and $M=$512, respectively, where the number of antennas at the BS is larger than the number of ISAC devices to guarantee the performance gain. The total power for sensing and communication is set as $P_{\max}=$0.1W and the noise power of ISAC devices is $\theta_c^2=-$80~dBm. In the computational cost, the basic price is set as $\lambda=0.1$. %The pathloss model is set as $PL_k$(dB)=148.1+37.6$\log_{10}l_k$(dB). 
We assume that there are 3 targets at the location $\bar{\theta}_1=-40^{\circ}$, $\bar{\theta}_2=0^{\circ}$, and $\bar{\theta}_3=40^{\circ}$, respectively. Therefore, the ideal beampattern  $P_d(\theta_l)$ is given by 
\begin{eqnarray}
P_d(\theta_l)=\left\{\begin{array}{ll}
		1, ~\bar{\theta}_p-\frac{\triangle_\theta}{2}\leq\theta_l\leq\bar{\theta}_p+\frac{\triangle_\theta}{2}, p=1,2,3,\\
		\\
		0,~\text{otherwise,}
\end{array}\right.		
\end{eqnarray}
where $\bar{\theta}_p$ represents the direction of the $p$th target, and $\triangle_\theta=10^{\circ}$. The direction grids $\{\theta_l\}_{l=1}^L$ are uniformly sampled from $-90^{\circ}$ to $90^{\circ}$ with a resolution of $1^{\circ}$.
\subsection{Performance Comparison of the Proposed Algorithm}\label{subsection_performance_comparison}
In this subsection, we compare the performance of the proposed waveform precoding design algorithm with the following existing algorithms. 
\begin{itemize}
	\item \emph{Baseline 1, Joint Precoding and Computation Resource Allocation~(JPCRA)~\cite{ding2022joint}:} The algorithm jointly optimizes the individual transmit precoding for radar, communication and computation resource allocation, and the problem is solved by adopting alternating iterative optimization problem.
	\item \emph{Baseline 2, Queue-aware Lyapunov Optimization~(QLO)~\cite{yang2020queue}:} The algorithm adopts Lyapunov optimization to solve the queue scheduling model of user data packets in the joint communication-radar system.
	\item \emph{Baseline 3, String-Pulling Method~(SPM)~\cite{zhou2022joint}}: The algorithm decomposes the sensing-and-transmission energy consumption problem and adopts the classical string-pulling method to solve the problem.
\end{itemize}

\begin{figure}[htbp]
	{	\centering \includegraphics[width=3.1in]{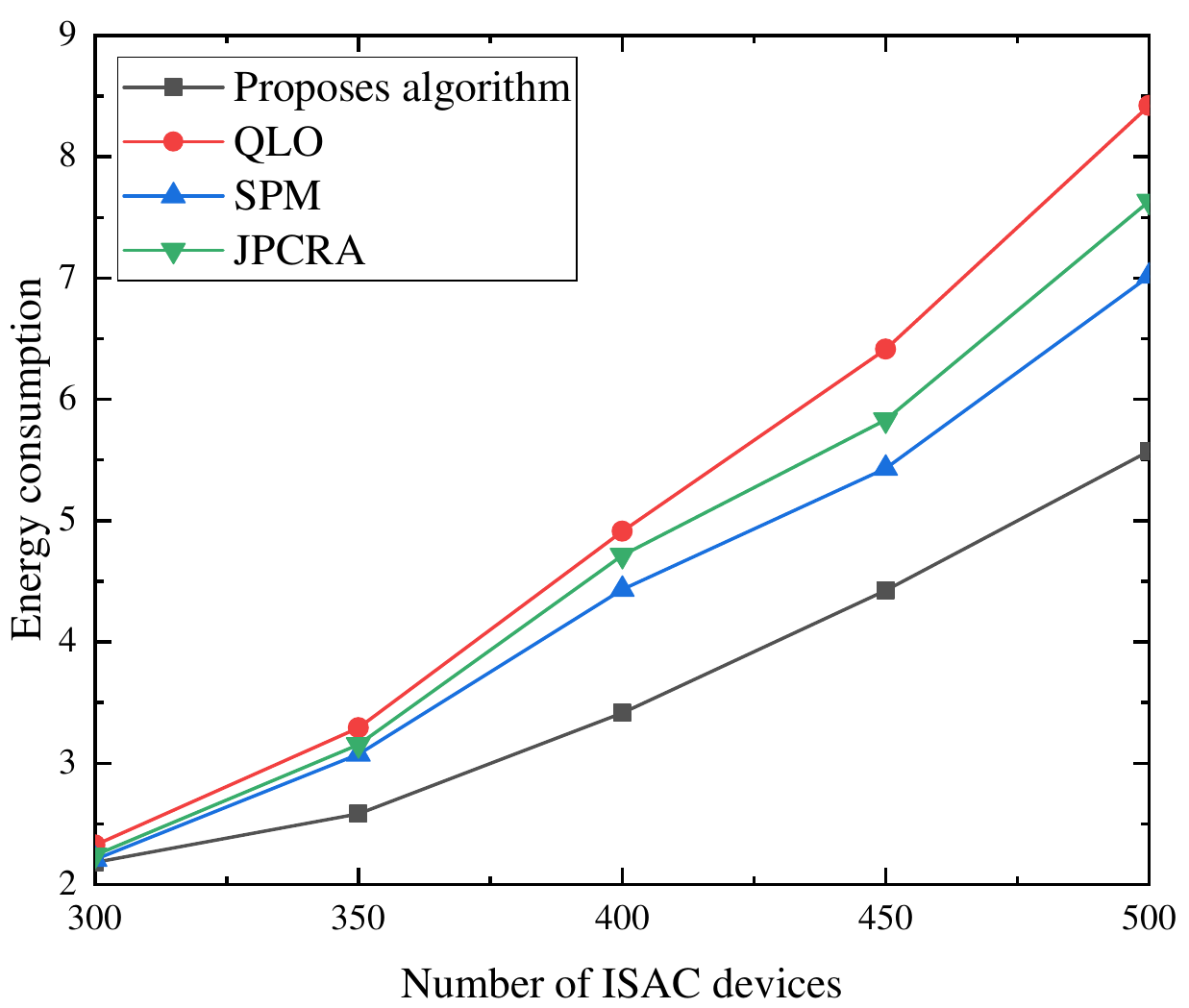}
		\caption{Energy consumption for different numbers of ISAC devices}
		\label{fig:energy_consumption_different_sensors}
	}
\end{figure}

Fig.~\ref{fig:energy_consumption_different_sensors} shows the energy consumption of the four algorithms versus the different number of ISAC devices. In the simulation, we consider the number of ISAC devices is smaller than the number of antennas at the BS to ensure the performance gain. It can be seen that energy consumption increases with the increased number of ISAC devices. This is because, with the increased number of ISAC devices, the interference among them will become more serious, which reduces communication efficiency and increases energy consumption. Compared with the other baselines, the proposed algorithm consumes less energy as the number of devices increases since it can handle the interference among ISAC devices effectively. When the number of ISAC devices is 500, the proposed algorithm reduces energy consumption by about 25\% compared with the SPM algorithm. 

\begin{figure}[htbp]
	{	\centering \includegraphics[width=3.1in]{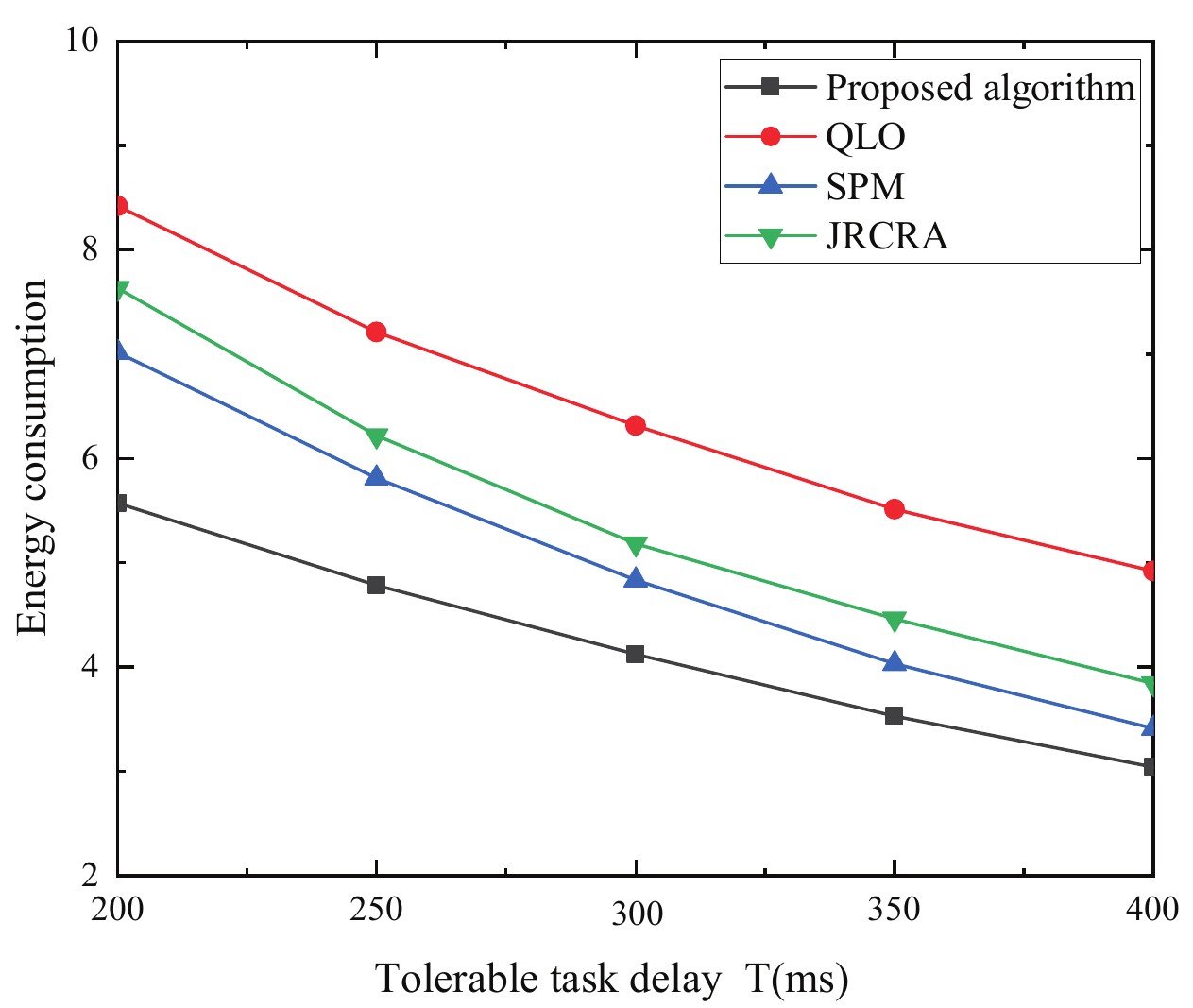}
		\caption{Energy consumption for different tolerable task delay}
		\label{fig:energy_consumption_different_tolerance}
	}
\end{figure}

Fig.~\ref{fig:energy_consumption_different_tolerance} shows the energy consumption of the four algorithms versus different tolerable task delays. It can be seen that the energy consumption for the four algorithms decreases with the increase of the tolerable task delay. The reason is that enlarging tolerable task delay can reduce the interference among ISAC devices and save energy consumption for sensing and communication. Compared with other baseline algorithms, the proposed algorithm consumes less energy with different tolerable time delays and reduces energy consumption by about 12\% when the tolerable delay is 400ms since it reduces the impact of interference among ISAC devices in the large-scale mobile crowd ISCC system. In addition, the gap between the proposed algorithm and other baselines decreases as the tolerable task delay increases, which is because the interference between ISAC devices decreases as the tolerable task delay increases.

\begin{figure}[htbp]
	{	\centering \includegraphics[width=3.1in]{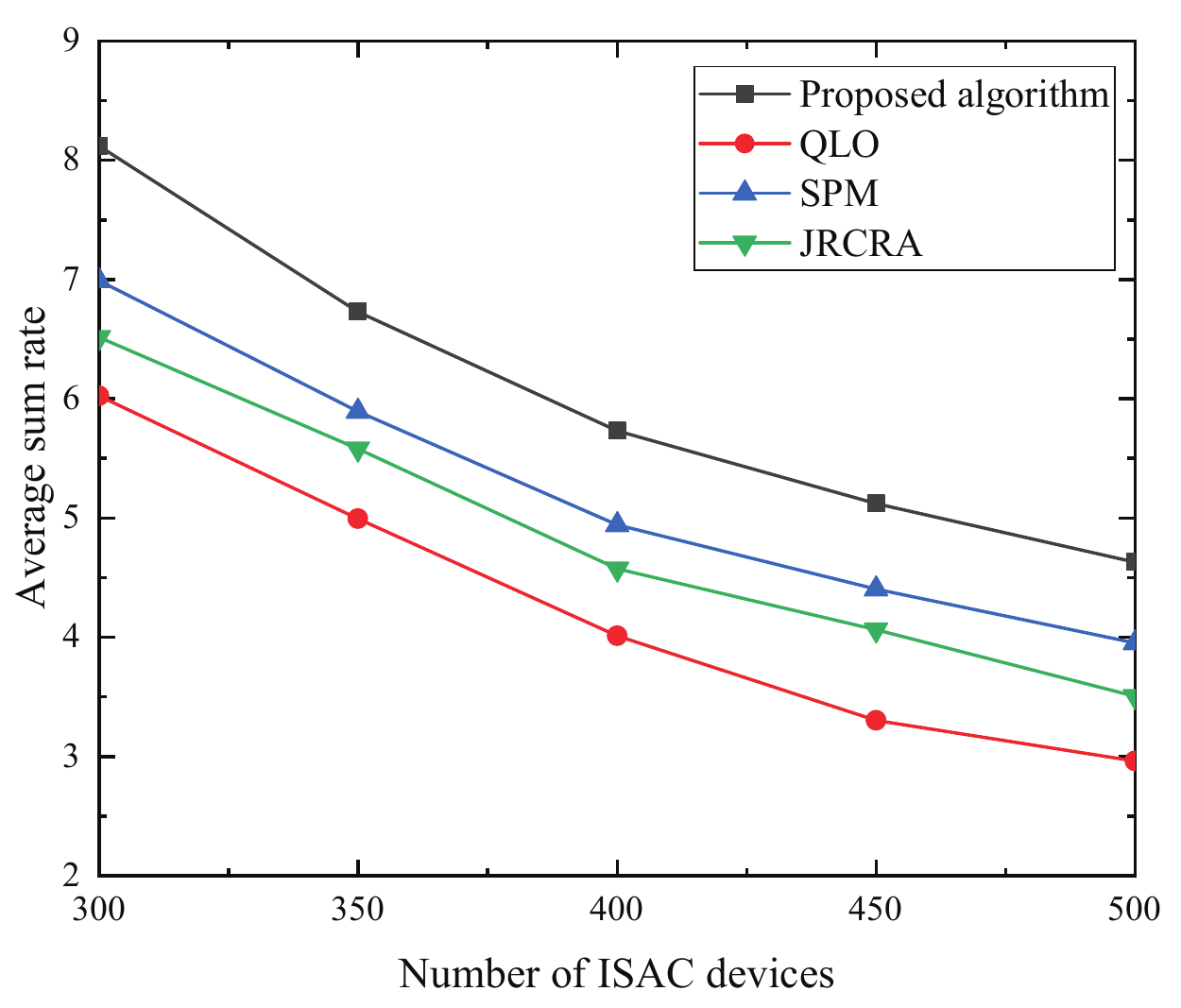}
		\caption{Average sum rate for different numbers of ISAC devices}
		\label{fig:average_sum_rate_different_sensors}
	}
\end{figure}

\begin{figure}[htbp]
	{	\centering \includegraphics[width=3.1in]{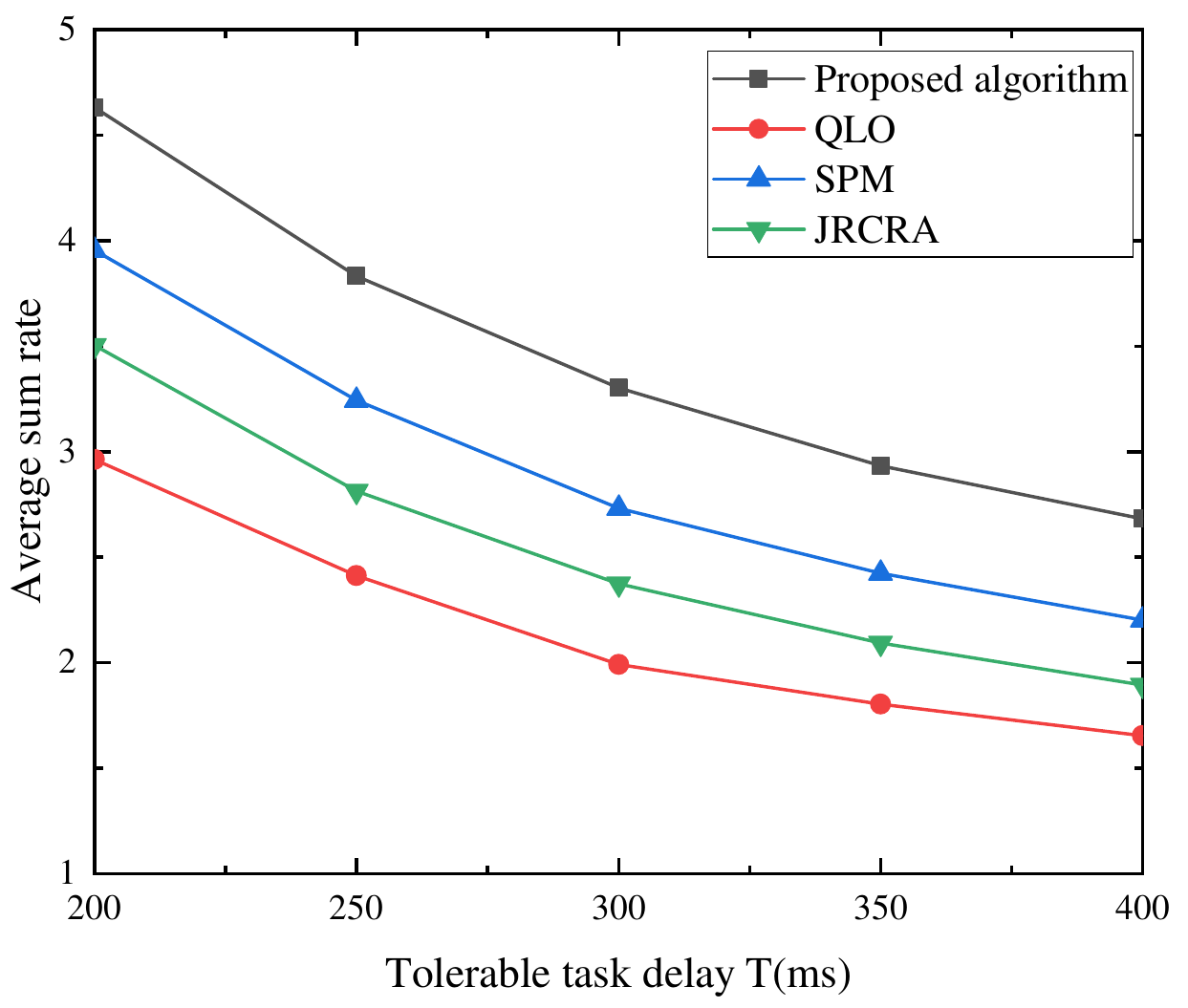}
		\caption{Average sum rate for different tolerable task delay}
		\label{fig:average_sum_rate_different_time_tolerance}
	}
\end{figure}

In addition, to show the communication performance of the proposed algorithm, we also compare the average sum rate of the proposed algorithm with other baselines in Figs.~\ref{fig:average_sum_rate_different_sensors} and~\ref{fig:average_sum_rate_different_time_tolerance}. Specifically, Fig.~\ref{fig:average_sum_rate_different_sensors} presents the average sum rate of the four algorithms versus the number of ISAC devices. It is obvious that the average sum rate of the four algorithms decreases with the increased number of ISAC devices and the proposed algorithm is superior to other baselines. This is because the increase in the number of ISAC devices leads to an increase in interference between ISAC devices, which affects communication efficiency, thus the corresponding average sum rate will decrease. Compared with other baselines, the proposed algorithm can effectively reduce the interference among ISAC devices, thus resulting in about a 15\% increase in transmission rate. Fig.~\ref{fig:average_sum_rate_different_time_tolerance} illustrates the average sum rate versus the tolerable task delay. It can be also seen that the average sum rate decreases with the increase in the tolerable task delay for the four algorithms. The reason is that increasing the tolerable task delay will lead to the data being transmitted over a longer period under the condition of constant total data volume. Similarly, the proposed algorithm outperforms other baselines for different tolerable task delays.

\subsection{Performance Analysis of the Proposed Algorithm}\label{subsection_performance_analysis}
\begin{figure}
	{	\centering \includegraphics[width=3.1in]{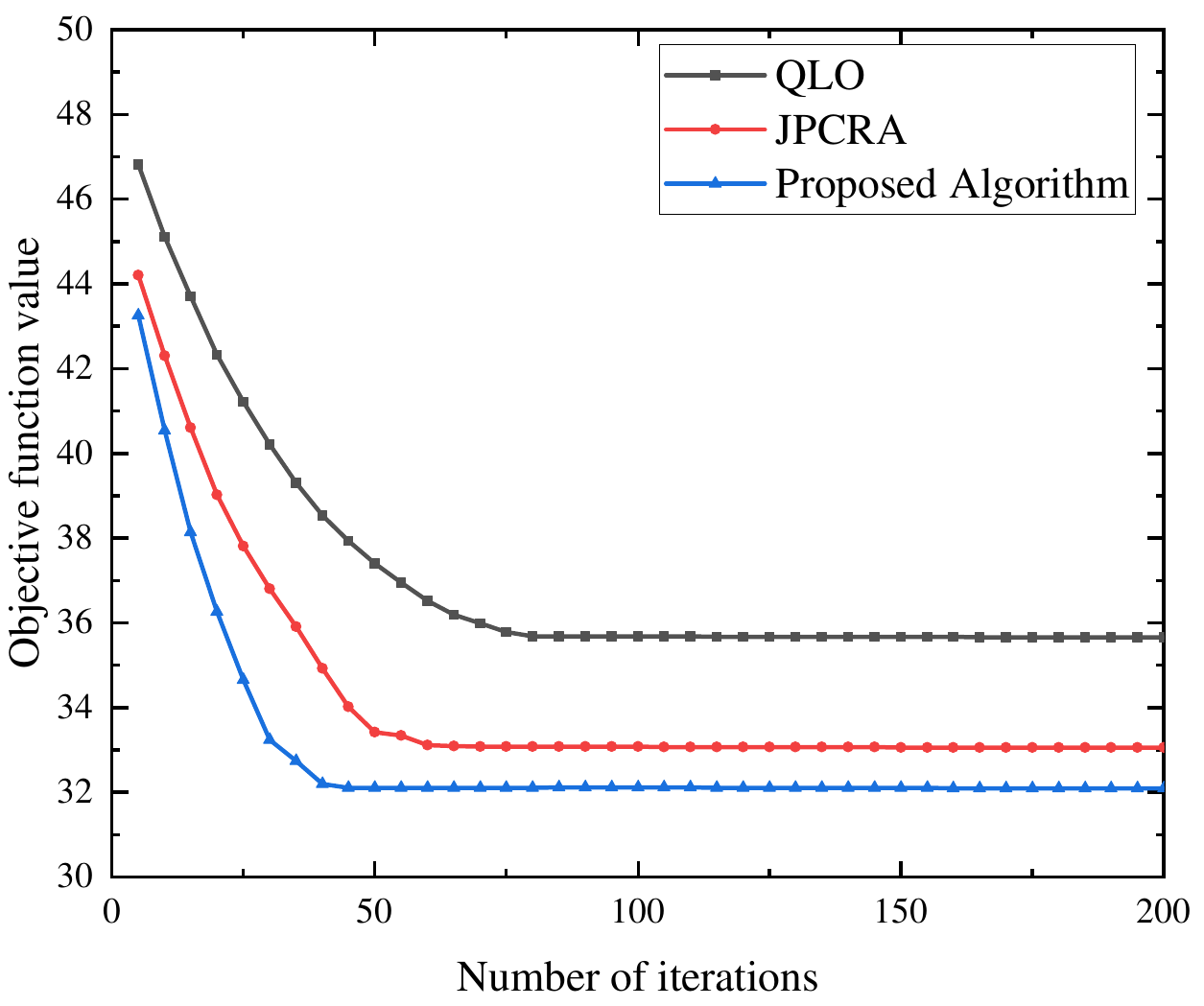} 		\caption{Convergence analysis for different algorithms }
		\label{fig:convergence}
	}
\end{figure}

In this subsection, the numerical results are presented to illustrate the convergence and effectiveness of the proposed waveform precoding design algorithm. At first, we compare the convergence of the proposed with other baselines in Fig.~\ref{fig:convergence}. Baseline 3 does not involve any iterations, so we compare the convergence performance with baseline 1 and baseline 2. As shown in the figure, the proposed algorithm achieves rapid convergence within 50 iterations, which is faster than baseline 1 and baseline 2. The reason is that the complex interference from other ISAC devices will affect the convergence performance of other baselines, which is different from the proposed algorithm.

\begin{figure}[htbp]
	{	\centering \includegraphics[width=3.1in]{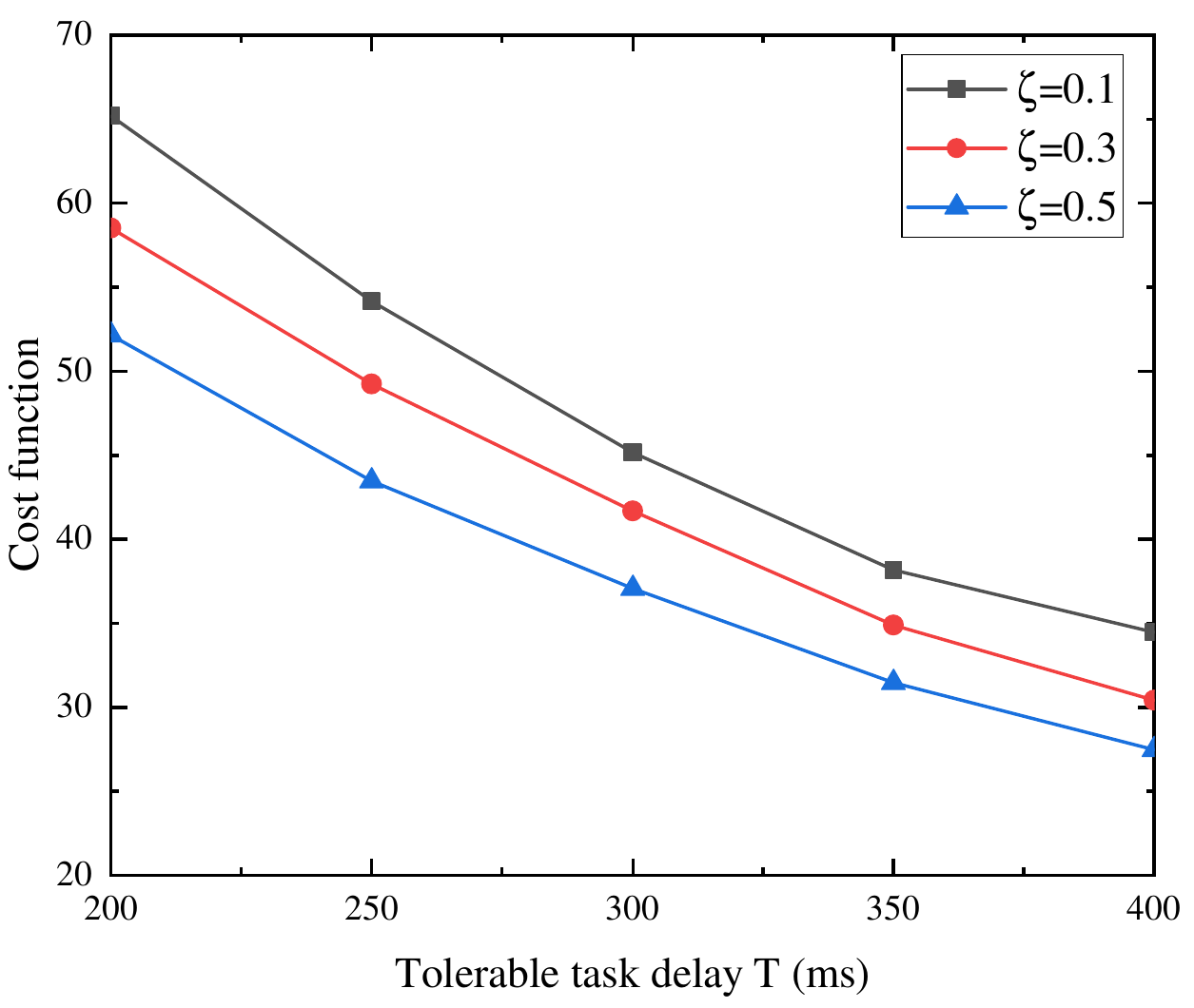}
		\caption{Cost function for different tolerable task delay}
		\label{fig:time_tolerance}
	}
\end{figure}

Fig.~\ref{fig:time_tolerance} shows the cost function versus the tolerable task delay $T$ under different thresholds~$\zeta$. It is obvious that as the tolerable task delay increases, the cost function decreases. This is because a longer tolerable task delay allows the ISAC devices to justify their waveform precoding design strategy to reduce interference among ISAC devices, increase transmission efficiency, and utilize computing resources at a lower cost. As a result, the cost function is reduced.  Furthermore, the results show that the cost function increases as the sensing accuracy improves, i.e., the value of $\zeta$ decreases. This can be attributed to the fact that higher accuracy requires ISAC devices to consume more energy to meet the sensing requirements, leading to an increase in the cost function, which shows that there is a trade-off between sensing accuracy and energy consumption.

\begin{figure}[htbp]
	{	\centering \includegraphics[width=3.1in]{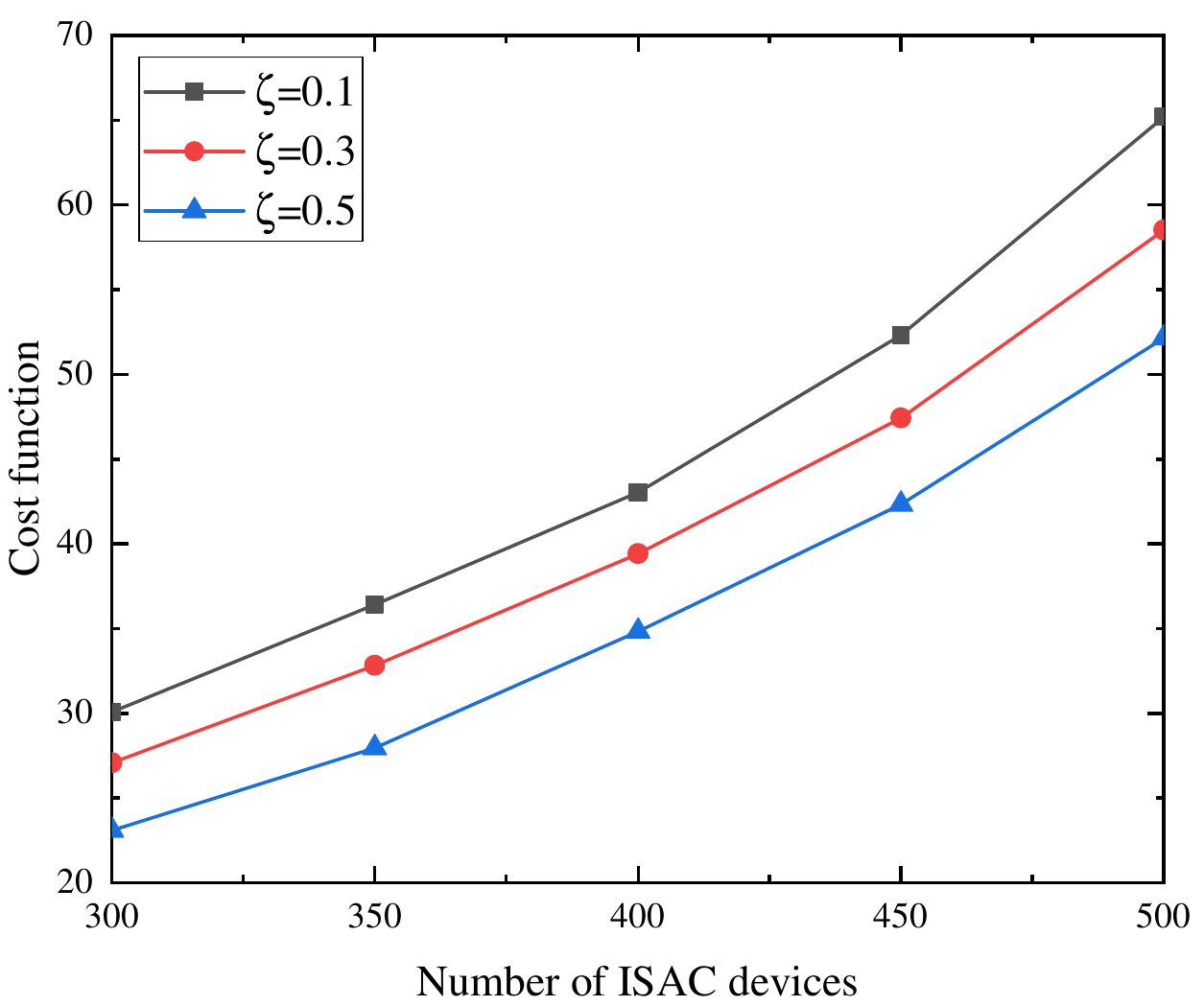}
		\caption{Cost function for different number of ISAC devices }
		\label{fig:number of sensor}
	}
\end{figure}

Fig.~\ref{fig:number of sensor} shows the cost function for different numbers of ISAC devices under different thresholds~$\zeta$. It can be seen that as the number of ISAC devices increases, the cost function also increases accordingly. This is because, with the increase in the number of ISAC devices, the interference among ISAC devices in the communication process becomes more severe, which seriously affects communication efficiency and consumes more energy. On the other hand, the data that needs to be processed increases with the increase in the number of ISAC devices, which increases the computational cost according to~\eqref{computation_cost_reformulation}. These dual increases will inevitably lead to an increase in the cost function.

\begin{figure}[htbp]
	{	\centering \includegraphics[width=3.1in]{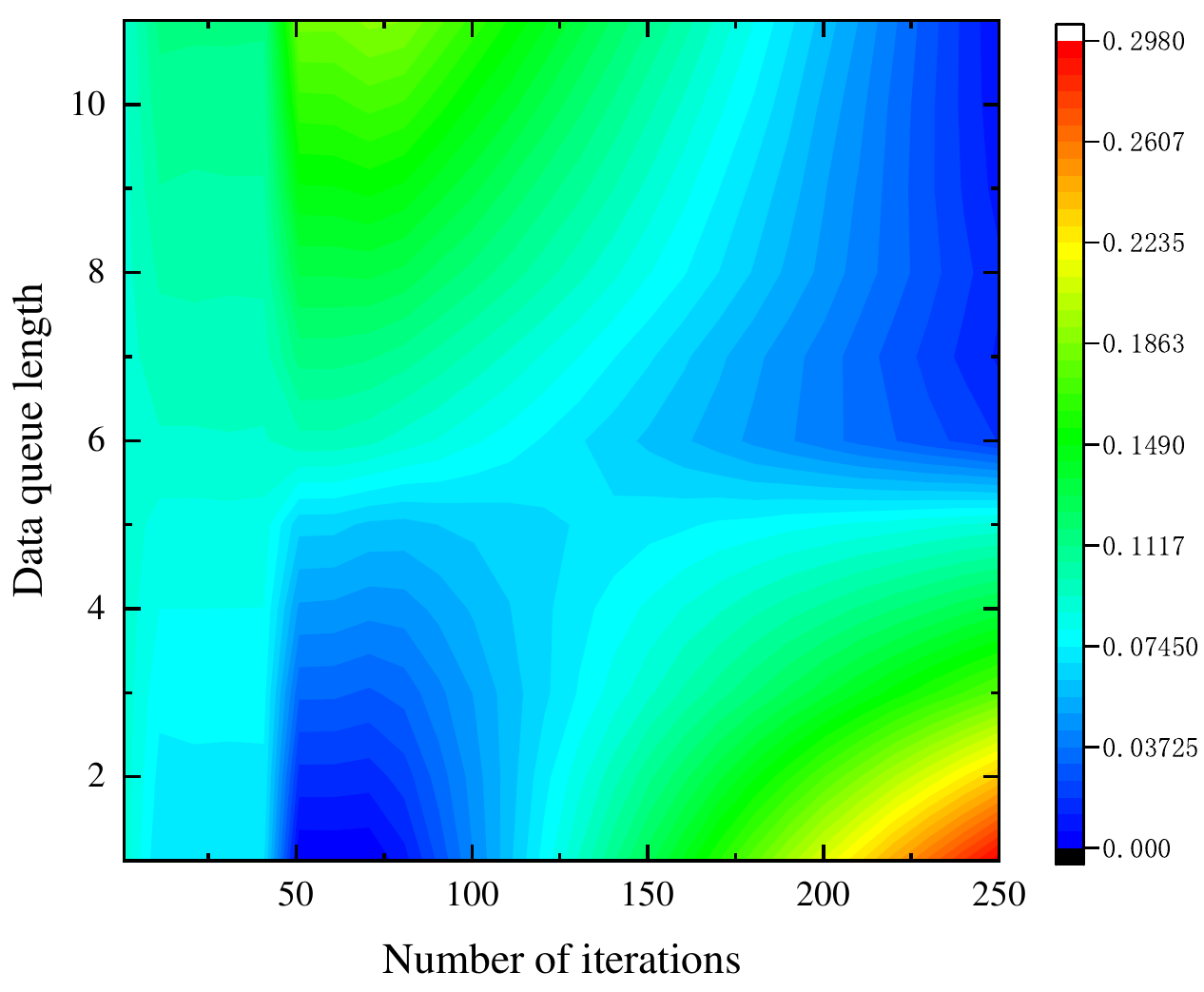}
		\caption{Evolution of the mean field term}
		\label{fig:mean_field_evoultion}
	}
\end{figure}

Further, Fig.~\ref{fig:mean_field_evoultion} shows the evolution process of the mean field term, i.e., $\rho(q,t)$, changing with the ISCC process. It can be seen from Fig.~\ref{fig:mean_field_evoultion} that the mean field term gradually changes from an initial uniform distribution to being concentrated in the area where the data queue is relatively higher, and finally being concentrated in the area where the data queue is relatively low in the ISCC processes. This is because as the data is collected, the data queue shows a corresponding increase, and then as data is continuously transmitted to the BS for computation, the corresponding data queue will decrease, finally being concentrated in the area where the data is zero. The result in Fig.~\ref{fig:mean_field_evoultion} also demonstrates the proposed algorithm can effectively process the data and ensure the delay performance of the system.

\section{Conclusion}
In this paper, we have studied the waveform precoding design problem in the large-scale mobile crowd ISCC system, where each ISAC device minimized its cost considering the influence from other ISAC devices. To solve the massive interactions among ISAC devices during the ISCC process, we first utilized the MFG method to transform the influence from other ISAC devices into the mean field term, which can effectively handle the interactions among large-scale ISAC devices. Further, the cross-layer joint optimization problem was modeled as a minimization problem of the energy consumption in sensing and communication processes, as well as the computational cost, while satisfying the sensing accuracy. Then, we derived the FPK equation to model the evolution of the system state and reformulated the minimization problem by utilizing the MFG method. Next, the reformulated problem was solved by using the G-prox PDHG algorithm with linear computational complexity. Finally, the simulation results showed that the proposed algorithm can reach convergence faster, and compared with other baseline algorithms, it can effectively reduce the interference among ISAC devices in the mobile crowd ISCC process, which further increases energy efficiency.

\appendices
\section{The Derivation of FPK Equation}
To derive the FPK equation~\cite{schulte2010adjoint}, we fist define $\phi(q)$ as a smooth and compactly supported function. Then, the integral $\int\rho(q,t)\phi(q,t)dq$ can be understood as the limit of the sum $\frac{1}{N}\sum_{i=1}^N\phi(q_i(t),t)$ in the continuum, where $q_i(t)$ is the state of ISAC device $i$ at time slot $t$. Applying the chain rule, we can differentiate the two parts w.r.t. $t$ as 
	\begin{eqnarray}
		\int\partial_t\rho(q,t)\phi(q,t)dq\approx\frac{1}{N}\sum_{i=1}^N\frac{\partial\phi(q_i(t),t)}{\partial q_i(t)}\frac{\partial q_i(t)}{\partial t}.
	\end{eqnarray}
	As the number of ISAC devices $N\rightarrow\infty$, we have 
	\begin{eqnarray}
		\int\partial_t\rho(q,t)\phi(q,t)dq=\int\frac{\partial\phi(q(t),t)}{\partial q(t)}\frac{\partial q(t)}{\partial t}\rho(q,t)dq.
	\end{eqnarray}
	Integration by parts, we have
	\begin{eqnarray}
		\int\big(\partial_t\rho(q,t)+\partial_q\rho(q,t)\partial_t q\big)\phi(q(t),t)dq.
	\end{eqnarray}
	which satisfies for arbitrary test function $\phi(q(t),t)$. Letting $\phi(q(t),t)=1$, we can obtain
	\begin{eqnarray}
		\partial_t\rho(q,t)+\partial_q(\rho(q,t)\partial_tq)=0,
	\end{eqnarray}
	which is exactly the FPK equation.

\section{Proof of Proposition 1 and Proposition 2}
To derive the gradient of the Lagrangian function w.r.t. $\rho$, we first transform the Lagrangian function based on the integration by parts, which is expressed as 
\begin{eqnarray}
&&L(\bm{W},\rho,\phi_1,\phi_2,\phi_3)\nonumber\\
&&=\int_0^T\int_{\mathcal{S}}\Big[\beta_1\|\bm{W}(q,t)\|_F^2+\beta_2\Phi(q,t)R(q,t)\Big]\rho(q,t)\nonumber\\
&&~~+\phi_1(q,t)c_1(q,t)+\phi_2(q,t)c_2(q,t)+\rho(q,t)\partial_t\phi_3(q,t)\nonumber\\
&&~~+\rho(q,t)\Gamma(q,t)\partial_q(\phi_3(q,t))dqdt+\int_{\mathcal{S}}\big(\rho(q,0)\phi_3(q,0)\nonumber\\
&&~~~~-\rho(q,T)\phi_3(q,T)\big)dq+\int_{\mathcal{S}} Cq(T)\rho(q,T)dq.
\end{eqnarray}
Thus, the gradient of $\mathcal{L}_\rho$ w.r.t. $\rho$ is calculated by
\begin{eqnarray}
&&\frac{\partial\mathcal{L}_{\rho}}{\partial\rho}=\beta_1\|\bm{W}(q,t)\|_F^2+\beta_2\Phi(q,t)R(q,t)+\partial_t\phi_3(q,t)\nonumber\\
&&~~+\Gamma(q,t)\partial_q(\phi_3(q,t))+\frac{1}{\tau}(\rho(q,t)-\rho^k(q,t)).
\end{eqnarray}
Similarly, the gradient of $\mathcal{L}_{\bm{W}}$ w.r.t. $\bm{W}$ is derived as 
\begin{eqnarray}
&&\frac{\partial\mathcal{L}_{\bm{W}}}{\partial\bm{W}}=\nonumber\\
&&\bigg(2\beta_1\bm{W}(q,t)+2\beta_2\frac{B\Phi(q,t)\bm{H}^H(t)\bm{H}(t)\bm{\Xi}(q,t)}{\ln2\arrowvert\bm{H}^H(t)\bm{\Xi}(q,t)\arrowvert^2}\bigg)\nonumber\\
&&\times\rho(q,t)+\frac{1}{2}\phi_1(q,t)(\bm{B}_1^k+(\bm{B}_1^k)^H+2\bm{B}_2^k+2(\bm{B}_2^k)^H)\nonumber\\
&&\times\bm{W}(q,t)+2\phi_2(q,t)\bm{W}(q,t)-\frac{2B\rho(q,t)}{\ln2(1+x_0)}\partial_q\phi_3(q,t)\nonumber\\
&&\times\frac{\bm{H}^H(t)\bm{H}(t)\bm{\Xi}(q,t)}{(\arrowvert\bm{n}\arrowvert^2+\sigma_c^2)}+\frac{1}{\tau}(\bm{W}(q,t)-\bm{W}^k(q,t)),
\end{eqnarray}
where $\bm{\Xi}(q,t)\overset{\triangle}{=}[\bm{0}_{M\times1}~\bm{w}_c(q,t)]$.\\
\balance
\bibliographystyle{IEEEtran}
\bibliography{myref}

\begin{thebibliography}{10}
\providecommand{\url}[1]{#1}
\csname url@samestyle\endcsname
\providecommand{\newblock}{\relax}
\providecommand{\bibinfo}[2]{#2}
\providecommand{\BIBentrySTDinterwordspacing}{\spaceskip=0pt\relax}
\providecommand{\BIBentryALTinterwordstretchfactor}{4}
\providecommand{\BIBentryALTinterwordspacing}{\spaceskip=\fontdimen2\font plus
\BIBentryALTinterwordstretchfactor\fontdimen3\font minus \fontdimen4\font\relax}
\providecommand{\BIBforeignlanguage}[2]{{%
\expandafter\ifx\csname l@#1\endcsname\relax
\typeout{** WARNING: IEEEtran.bst: No hyphenation pattern has been}%
\typeout{** loaded for the language `#1'. Using the pattern for}%
\typeout{** the default language instead.}%
\else
\language=\csname l@#1\endcsname
\fi
#2}}
\providecommand{\BIBdecl}{\relax}
\BIBdecl

\bibitem{wang2023GC}
D.~Wang, C.~Huang, J.~He, C.~Zhu, W.~Wang, X.~Chen, Z.~Zhang, Z.~Han, and M.~Debbah, ``Waveform precoding design for mobile crowd {ISCC} system using mean field game,'' in \emph{Proc. IEEE Global Commun. Conf. Workshops.~(Globecom WKSHPS)}.\hskip 1em plus 0.5em minus 0.4em\relax IEEE, Dec. 2023.

\bibitem{jamshed2022challenges}
M.~A. Jamshed, K.~Ali, Q.~H. Abbasi, M.~A. Imran, and M.~Ur-Rehman, ``Challenges, applications and future of wireless sensors in {Internet of Things}: A review,'' \emph{IEEE Sensors J.}, vol.~22, no.~6, pp. 5482--5494, Jan. 2022.

\bibitem{capponi2019survey}
A.~Capponi, C.~Fiandrino, B.~Kantarci, L.~Foschini, D.~Kliazovich, and P.~Bouvry, ``A survey on mobile crowdsensing systems: Challenges, solutions, and opportunities,'' \emph{IEEE Commun. Surveys \& Tuts.}, vol.~21, no.~3, pp. 2419--2465, 3rd Quart. 2019.

\bibitem{zhu2020deep}
X.~Zhu, Y.~Luo, A.~Liu, W.~Tang, and M.~Z.~A. Bhuiyan, ``A deep learning-based mobile crowdsensing scheme by predicting vehicle mobility,'' \emph{IEEE Trans. Intell. Transp. Syst.}, vol.~22, no.~7, pp. 4648--4659, Jul. 2021.

\bibitem{zhang2022achieving}
Y.~Zhang, Z.~Ying, and C.~P. Chen, ``Achieving privacy-preserving multitask allocation for mobile crowdsensing,'' \emph{IEEE Internet Things J.}, vol.~9, no.~18, pp. 16\,795--16\,806, Sep. 2022.

\bibitem{xiong2020edge}
J.~Xiong, R.~Bi, M.~Zhao, J.~Guo, and Q.~Yang, ``Edge-assisted privacy-preserving raw data sharing framework for connected autonomous vehicles,'' \emph{IEEE Wireless Commun.}, vol.~27, no.~3, pp. 24--30, Jun. 2020.

\bibitem{liu2019data}
Y.~Liu, L.~Kong, and G.~Chen, ``Data-oriented mobile crowdsensing: A comprehensive survey,'' \emph{IEEE Commun. Surveys \& Tuts.}, vol.~21, no.~3, pp. 2849--2885, 3rd Quart. 2019.

\bibitem{luong2021radio}
N.~C. Luong, X.~Lu, D.~T. Hoang, D.~Niyato, and D.~I. Kim, ``Radio resource management in joint radar and communication: A comprehensive survey,'' \emph{IEEE Commun. Surveys \& Tuts.}, vol.~23, no.~2, pp. 780--814, 2nd Quart. 2021.

\bibitem{wang2023delay}
D.~Wang, W.~Wang, H.~Gao, Z.~Zhang, and Z.~Han, ``Delay-optimal computation offloading in large-scale multi-access edge computing using mean field game,'' \emph{IEEE Trans. Wireless Commun.}, vol.~99, no.~99, pp. 1--1, 2023.

\bibitem{he2023integrated}
Y.~He, G.~Yu, Y.~Cai, and H.~Luo, ``Integrated sensing, computation, and communication: system framework and performance optimization,'' \emph{IEEE Trans. Wireless Commun.}, vol.~99, no.~99, pp. 1--1, 2023.

\bibitem{wen2023task2}
D.~Wen, X.~Jiao, P.~Liu, G.~Zhu, Y.~Shi, and K.~Huang, ``Task-oriented over-the-air computation for multi-device edge ai,'' \emph{IEEE Trans. Wireless Commun.}, vol.~99, no.~99, pp. 1--1, 2023.

\bibitem{liu2022integrated}
F.~Liu, Y.~Cui, C.~Masouros, J.~Xu, T.~X. Han, Y.~C. Eldar, and S.~Buzzi, ``Integrated sensing and communications: Towards dual-functional wireless networks for {6G} and beyond,'' \emph{IEEE J. Sel. Areas Commun.}, vol.~40, no.~6, pp. 1728--1767, Mar. 2022.

\bibitem{cui2021integrating}
Y.~Cui, F.~Liu, X.~Jing, and J.~Mu, ``Integrating sensing and communications for ubiquitous {IoT:} applications, trends, and challenges,'' \emph{IEEE Netw.}, vol.~35, no.~5, pp. 158--167, Sep.-Oct. 2021.

\bibitem{du2022integrated}
Z.~Du, F.~Liu, W.~Yuan, C.~Masouros, Z.~Zhang, S.~Xia, and G.~Caire, ``Integrated sensing and communications for {V2I} networks: Dynamic predictive beamforming for extended vehicle targets,'' \emph{IEEE Trans. Wireless Commun.}, vol.~22, no.~6, pp. 3612--3627, Jun. 2023.

\bibitem{hua2023optimal}
H.~Hua, J.~Xu, and T.~X. Han, ``Optimal transmit beamforming for integrated sensing and communication,'' \emph{IEEE Trans. Veh. Technol.}, vol.~99, no.~99, pp. 1--1, 2023.

\bibitem{tong2021joint}
X.~Tong, Z.~Zhang, J.~Wang, C.~Huang, and M.~Debbah, ``Joint multi-user communication and sensing exploiting both signal and environment sparsity,'' \emph{IEEE J. Sel. Topics Signal Process.}, vol.~15, no.~6, pp. 1409--1422, Sep. 2021.

\bibitem{gan2022near}
X.~Gan, C.~Huang, Z.~Yang, C.~Zhong, and Z.~Zhang, ``Near-field localization for holographic ris assisted mmwave systems,'' \emph{IEEE Commun. Lett.}, vol.~27, no.~1, pp. 140--144, Jan. 2023.

\bibitem{wen2023task}
D.~Wen, P.~Liu, G.~Zhu, Y.~Shi, J.~Xu, Y.~C. Eldar, and S.~Cui, ``Task-oriented sensing, computation, and communication integration for multi-device edge {AI},'' \emph{IEEE Trans. Wireless Commun.}, vol.~99, no.~99, pp. 1--1, 2023.

\bibitem{qi2022integrating}
Q.~Qi, X.~Chen, A.~Khalili, C.~Zhong, Z.~Zhang, and D.~W.~K. Ng, ``Integrating sensing, computing, and communication in {6G} wireless networks: Design and optimization,'' \emph{IEEE Trans. Commun.}, vol.~70, no.~9, pp. 6212--6227, Sep. 2022.

\bibitem{ding2022joint}
C.~Ding, J.-B. Wang, H.~Zhang, M.~Lin, and G.~Y. Li, ``Joint {MIMO} precoding and computation resource allocation for dual-function radar and communication systems with mobile edge computing,'' \emph{IEEE J. Sel. Areas Commun.}, vol.~40, no.~7, pp. 2085--2102, Jul. 2022.

\bibitem{li2023integrated}
X.~Li, F.~Liu, Z.~Zhou, G.~Zhu, S.~Wang, K.~Huang, and Y.~Gong, ``Integrated sensing, communication, and computation over-the-air: {MIMO} beamforming design,'' \emph{IEEE Trans. Wireless Commun.}, vol.~99, no.~99, pp. 1--1, 2023.

\bibitem{huang2022integrated}
N.~Huang, T.~Wang, Y.~Wu, Q.~Wu, and T.~Q. Quek, ``Integrated sensing and communication assisted mobile edge computing: An energy-efficient design via intelligent reflecting surface,'' \emph{IEEE Wireless Commun. Lett.}, vol.~11, no.~10, pp. 2085--2089, Oct. 2022.

\bibitem{li2023quality}
Z.~Li, Z.~Li, and W.~Zhang, ``Quality-aware task allocation for mobile crowd sensing based on edge computing,'' \emph{Electronics}, vol.~12, no.~4, p. 960, Feb. 2023.

\bibitem{xiong2019task}
J.~Xiong, X.~Chen, Q.~Yang, L.~Chen, and Z.~Yao, ``A task-oriented user selection incentive mechanism in edge-aided mobile crowdsensing,'' \emph{IEEE Trans. Netw. Sci. Eng.}, vol.~7, no.~4, pp. 2347--2360, Oct.-Dec. 2020.

\bibitem{li2022joint}
X.~Li, G.~Feng, Y.~Liu, S.~Qin, and Z.~Zhang, ``Joint sensing, communication and computation in mobile crowdsensing enabled edge networks,'' \emph{IEEE Trans. Wireless Commun.}, vol.~22, no.~4, pp. 2818--2832, Apr. 2023.

\bibitem{zhou2022joint}
Z.~Zhou, X.~Li, C.~You, K.~Huang, and Y.~Gong, ``Joint sensing and communication-rate control for energy efficient mobile crowd sensing,'' \emph{IEEE Trans. Wireless Commun.}, vol.~22, no.~2, pp. 1314--1327, Feb. 2023.

\bibitem{cai2021cooperative}
T.~Cai, Z.~Yang, Y.~Chen, W.~Chen, Z.~Zheng, Y.~Yu, and H.-N. Dai, ``Cooperative data sensing and computation offloading in {UAV-assisted} crowdsensing with multi-agent deep reinforcement learning,'' \emph{IEEE Trans. Netw. Sci. Eng.}, vol.~9, no.~5, pp. 3197--3211, Sep.-Oct. 2022.

\bibitem{kang2021task}
Y.~Kang, S.~Liu, H.~Zhang, Z.~Han, S.~Osher, and H.~V. Poor, ``Task selection and collision-free route planning for mobile crowdsensing using multi-population mean-field games,'' \emph{IEEE Trans. Green Commun. Netw.}, vol.~5, no.~4, pp. 1947--1960, Dec. 2021.

\bibitem{xing2023joint}
Z.~Xing, R.~Wang, and X.~Yuan, ``Joint active and passive beamforming design for reconfigurable intelligent surface enabled integrated sensing and communication,'' \emph{IEEE Trans. Commun.}, vol.~71, no.~4, pp. 2457--2474, Apr. 2023.

\bibitem{xiong2023fundamental}
Y.~Xiong, F.~Liu, Y.~Cui, W.~Yuan, T.~X. Han, and G.~Caire, ``On the fundamental tradeoff of integrated sensing and communications under {Gaussian} channels,'' \emph{IEEE Trans. Inf. Theory}, vol.~99, no.~99, pp. 1--1, 2023.

\bibitem{an2023fundamental}
J.~An, H.~Li, D.~W.~K. Ng, and C.~Yuen, ``Fundamental detection probability vs. achievable rate tradeoff in integrated sensing and communication systems,'' \emph{IEEE Trans. Wireless Commun.}, vol.~99, no.~99, pp. 1--1, 2023.

\bibitem{gao2022energyed}
H.~Gao, W.~Lee, Y.~Kang, W.~Li, Z.~Han, S.~Osher, and H.~V. Poor, ``Energy-efficient velocity control for massive numbers of {UAV}s: A mean field game approach,'' \emph{IEEE Trans. Vehi. Technol.}, vol.~71, no.~6, pp. 6266--6278, Jun. 2022.

\bibitem{yang2017mean}
C.~Yang, J.~Li, M.~Sheng, A.~Anpalagan, and J.~Xiao, ``Mean field game-theoretic framework for interference and energy-aware control in {5G} ultra-dense networks,'' \emph{IEEE Wireless Commun.}, vol.~25, no.~1, pp. 114--121, Feb. 2018.

\bibitem{khawar2015target}
A.~Khawar, A.~Abdelhadi, and C.~Clancy, ``Target detection performance of spectrum sharing {MIMO} radars,'' \emph{IEEE Sensors J.}, vol.~15, no.~9, pp. 4928--4940, Sep. 2015.

\bibitem{liu2018mimo}
F.~Liu, C.~Masouros, A.~Li, T.~Ratnarajah, and J.~Zhou, ``{MIMO} radar and cellular coexistence: A power-efficient approach enabled by interference exploitation,'' \emph{IEEE Trans. Signal Process.}, vol.~66, no.~14, pp. 3681--3695, Jul. 2018.

\bibitem{bekkerman2006target}
I.~Bekkerman and J.~Tabrikian, ``Target detection and localization using {MIMO} radars and sonars,'' \emph{IEEE Trans. Signal Process.}, vol.~54, no.~10, pp. 3873--3883, Oct. 2006.

\bibitem{farina2006introduction}
A.~Farina, ``Introduction to radar signal and data processing: The opportunity,'' Selex Sistemi Integrati Rome (Italy), Tech. Rep., 2006.

\bibitem{liang2019multiuser}
Z.~Liang, Y.~Liu, T.-M. Lok, and K.~Huang, ``Multiuser computation offloading and downloading for edge computing with virtualization,'' \emph{IEEE Trans. Wireless Commun.}, vol.~18, no.~9, pp. 4298--4311, Sep. 2019.

\bibitem{zhang2017fog}
H.~Zhang, Y.~Qiu, X.~Chu, K.~Long, and V.~C. Leung, ``Fog radio access networks: Mobility management, interference mitigation, and resource optimization,'' \emph{IEEE Wireless Commun.}, vol.~24, no.~6, pp. 120--127, Dec. 2017.

\bibitem{ku20175g}
Y.-J. Ku, D.-Y. Lin, C.-F. Lee, P.-J. Hsieh, H.-Y. Wei, C.-T. Chou, and A.-C. Pang, ``{5G} radio access network design with the fog paradigm: Confluence of communications and computing,'' \emph{IEEE Commun. Mag.}, vol.~55, no.~4, pp. 46--52, Dec. 2017.

\bibitem{zheng2021dynamic}
R.~Zheng, H.~Wang, M.~De~Mari, M.~Cui, X.~Chu, and T.~Q. Quek, ``Dynamic computation offloading in ultra-dense networks based on mean field games,'' \emph{IEEE Trans. Wireless Commun.}, vol.~20, no.~10, pp. 6551--6565, Oct. 2021.

\bibitem{huang2019reconfigurable}
C.~Huang, A.~Zappone, G.~C. Alexandropoulos, M.~Debbah, and C.~Yuen, ``Reconfigurable intelligent surfaces for energy efficiency in wireless communication,'' \emph{IEEE Trans. Wireless Commun.}, vol.~18, no.~8, pp. 4157--4170, Aug. 2019.

\bibitem{rihan2018optimum}
M.~Rihan and L.~Huang, ``Optimum co-design of spectrum sharing between {MIMO} radar and {MIMO} communication systems: An interference alignment approach,'' \emph{IEEE Trans. Vehi. Technol.}, vol.~67, no.~12, pp. 11\,667--11\,680, Dec. 2018.

\bibitem{luo2022joint}
H.~Luo, R.~Liu, M.~Li, Y.~Liu, and Q.~Liu, ``Joint beamforming design for {RIS}-assisted integrated sensing and communication systems,'' \emph{IEEE Trans. Veh. Technol.}, vol.~71, no.~12, pp. 13\,393--13\,397, Dec. 2022.

\bibitem{wang2022distributed}
D.~Wang, W.~Wang, Y.~Kang, and Z.~Han, ``Distributed data offloading in ultra-dense {LEO} satellite networks: A {Stackelberg} mean-field game approach,'' \emph{IEEE J. Sel. Topics Signal Process.}, vol.~17, no.~1, pp. 112--127, Jan. 2023.

\bibitem{banez2021mean}
R.~A. Banez, L.~Li, C.~Yang, and Z.~Han, \emph{Mean Field Game and Its Applications in Wireless Networks}.\hskip 1em plus 0.5em minus 0.4em\relax Springer, 2021.

\bibitem{wang2021delay}
D.~Wang, W.~Wang, Z.~Han, and Z.~Zhang, ``Delay optimal random access with heterogeneous device capabilities in energy harvesting networks using mean field game,'' \emph{IEEE Trans. Wireless Commun.}, vol.~20, no.~9, pp. 5543--5557, Sep. 2021.

\bibitem{lasry2007mean}
J.-M. Lasry and P.-L. Lions, ``Mean field games,'' \emph{Japanese journal of mathematics}, vol.~2, no.~1, pp. 229--260, 2007.

\bibitem{kang2020joint}
Y.~Kang, S.~Liu, H.~Zhang, W.~Li, Z.~Han, S.~Osher, and H.~V. Poor, ``Joint sensing task assignment and collision-free trajectory optimization for mobile vehicle networks using mean-field games,'' \emph{IEEE Internet Things J.}, vol.~8, no.~10, pp. 8488--8503, May 2021.

\bibitem{schlegel2015trellis}
C.~B. Schlegel and L.~C. Perez, \emph{Trellis and turbo coding: iterative and graph-based error control coding}.\hskip 1em plus 0.5em minus 0.4em\relax John Wiley \& Sons, 2015.

\bibitem{yang2020queue}
H.~Yang, Z.~Wei, Z.~Feng, C.~Qiu, Z.~Fang, X.~Chen, and P.~Zhang, ``Queue-aware dynamic resource allocation for the joint communication-radar system,'' \emph{IEEE Trans. Veh. Technol.}, vol.~70, no.~1, pp. 754--767, Jan. 2020.

\bibitem{schulte2010adjoint}
J.~M., Schulte, ``Adjoint methods for {Hamilton-Jacobi-Bellman} equations,'' \emph{Diploma Thesis}, Uinversity of Munster, Germany, Nov. 2010.

\end{thebibliography}

\end{document}